\documentclass{imsart}
\RequirePackage{natbib}
\usepackage{amsmath,amssymb,amsthm}
\usepackage{graphics,epsfig}
\usepackage{hyperref}
\usepackage{natbib}
\usepackage{color}
\usepackage{graphicx}
\usepackage{caption}
\usepackage{subcaption}
\usepackage{float}
\usepackage{dsfont}
\usepackage{mathrsfs}
\usepackage{multirow}
\usepackage{comment}
\usepackage{bm}

\def \qq {\boldsymbol{q}}

\def \ra {\rightarrow}

\def \E {\mathbb{E}}

\def \a {\alpha}
\def \be {\beta}

\newtheorem{example}{\bf Example}

\newtheorem{definition}{\bf Definition}
\newtheorem{defn}[definition]{\bf Definition}

	\newtheorem{theorem}{\bf Theorem}
	
	\newtheorem{prop}{\bf Proposition}
	\newtheorem{lem}[theorem]{\bf Lemma}
	\newtheorem{as}{\bf Assumption}

\renewcommand{\epsilon}{\varepsilon}

\begin{document}

\begin{frontmatter}

\title{Optimal Estimation of Brownian Penalized Regression Coefficients}
\runtitle{Regression Coefficients}

\begin{aug}

\author[A]{\fnms{Paramahansa} 
	\snm{Pramanik}
	\ead[label=e1]{ppramanik@southalabama.edu}}
\and
\author[B]{\fnms{Alan M.} 
	\snm{Polansky}
	\ead[label=e2]{polansky@niu.edu}}
	
\runauthor{P. Pramanik and A. M. Polansky}

\affiliation[A]{University of South Alabama}

\affiliation[B]{Northern Illinois University}

\address[A]{Department of Mathematics and Statistics\\ University of South Alabama\\ Mobile, AL 36688 USA.}

\address[B]{Department of Statistics and Actuarial Science \\ Northern Illinois University\\
	DeKalb, IL 60115 USA.}
\end{aug}
  
\begin{abstract}
In this paper we introduce  a new methodology to determine an optimal coefficient of penalized functional regression. We assume the dependent, independent variables and the regression coefficients are functions of time and error dynamics follow a stochastic differential equation. First we construct our objective function as a time dependent residual sum of square and then minimize it with respect to regression coefficients subject to different error dynamics such as LASSO, group LASSO, fused LASSO and cubic smoothing spline. Then we use Feynman-type path integral approach to determine a Schr\"odinger-type equation which have the entire information of the system. Using first order conditions with respect to these coefficients give us a closed form solution of them.
\end{abstract}



\end{frontmatter}

\section{Introduction}
Functional regression has been popular in recent times \citep{ramsay2004,ramsay2007}. Different penalizations such as least absolute shrinkage and selection operator (LASSO), ridge regression, standard $L^p$-norm, elastic net regression, Group LASSO, fused LASSO, bridge regression and different types of splines have been used in statistical literature for variable selection. Following \cite{huang2021} we know, penalized regression has been popularized after publication of \cite{eilers1996} and \cite{ruppert2003}. A mean squared error
of penalized spline estimators under a white noise model was obtained in \cite{hall2005}. Furthermore, different approximation of penalized spline estimators have been done in \cite{li2008}, \cite{wang2011}, \cite{schwarz2016} and \cite{lai2013}. These works have been used in closed-form expressions of
 penalized spline estimators which are only available in the regression setting where all the variables are time independent. When such expressions are not available in other estimation contexts, such as estimation of density functions or conditional quantile functions, \cite{huang2021} or furthermore, when the penalization function is itself a stochastic differential equation, existing
 asymptotic approaches extended. Then we need a path integral approach to determine regression coefficients in Euclidean field \citep{pramanik2020optimization,pramanik2020,pramanik2020motivation,pramanik2021,pramanik2021a} and for generalized tensor field \citep{pramanik2019}.
 
 In this paper we provide a dynamic framework of a time dependent  residual sum of square and minimize it with respect to regression coefficients where coefficient dynamics follow a stochastic differential equation. We construct a quantum Lagrangian for equal in length small time interval with respect to a positive penalization parameter and use a Feynman-type path integral approach to determine a Schr\"odinger type equation \citep{pramanik2016,hua2019,pramanik2020optimization,pramanik2021,polansky2021} and optimal values of the regression coefficients are the first order condition of it \citep{baaquie2007, feynman1949} and \cite{fujiwara2017}. As at the beginning of a new time interval we do not have any prior knowledge about the future, a conditional expectation until that initial time point of the residual sum of squares is used as our objective function. In examples we show closed form of the regression coefficients under different penalizations. Traditional literature of this type regression does not consider diffusion part of the process. Hence, we cannot see the bigger picture of it and  more generalization towards Brownian motion is needed. 

 Before constructing the quantum Lagrangian for small time intervals and path integral of the system we showed those two integrals exist under certain assumptions, which will be discussed in the next section. Main motivation of using Feynman path integral approach is it considers all possible paths between two time points and eliminates the extremes by Lebesgue-Riemann lemma to determine the minimized action locally. Furthermore, this approach gives solution for more generalized system of equations where Pontryagin's optimal principle fails \citep{baaquie2007, bellman1966} and \cite{yeung2006}. 

\section{Preliminaries}
Consider a sample of $N$ time dependent cases each of which consists of $J$ covariates  such that for an observation $i$ we have the following regression model
\begin{equation*}
Y_i(s)=\sum_{j'=1}^J\beta_{j'}(s)X_{ij'}(s)+{\bf U}_{i}(s),
\end{equation*}
where $\be_{j'}(s)\in \bm\be(s)\in\mathbb{R}^{J}$ for all $j'=1,...,J$, $Y_i(s)\in\mathbb{R}^{N}$ is $i^{th}$ outcome and $X_{ij'}(s)\in\mathbb{R}^{N\times J}$ is $i^{th}$ independent variable corresponding to deterministic $\be_{j'}(s)$ coefficient, with $i=1,...,N$ and time $s\in[0,T]$ and, the error term ${\bf U}_i\in{\bf U}\in\mathbb{R}^N$ is assumed to be a stochastic process expressed by the stochastic differential Equation (\ref{0}) below.

Therefore, to obtain an optimal regression coefficient the objective is to minimize time dependent residual sum of square (RSS)
\begin{equation*}
\overline{\mathbf{X}}_O(s,\bm\be,\mathbf{X})= \sum_{i=1}^{N}\left[Y_i(s)-\sum_{j'=1}^J\beta_{j'}(s)X_{ij'}(s)\right]^2,
\end{equation*}
 with respect to $\be_{j'}(s)\in \bm\be(s)\in\mathbb{R}^{J}$  Furthermore, we assume the $N$-dimensional error vector $\bf U(s)$ follows a stochastic  differential equation,
\begin{equation}\label{0}
d\mathbf{\bf U}(s)=\bm{\mu}[s,\bm\be(s),\mathbf{X}(s)]ds+
\bm{\sigma}[s,\bm\be(s),\mathbf{X}(s)]d\mathbf{B}(s),
\end{equation}
where $\bm{\mu}[s,\bm\be(s),\mathbf{X}(s)]$ is a $N\times 1$-dimensional drift vector, $\bm{\sigma}[s,\bm\be(s),\mathbf{X}(s)]$ a $N\times p$-dimensional diffusion matrix and $\mathbf{B}(s)$ is a $p$-dimensional Brownian motion. The mappings of $\bm\mu[s,\bm\be(s),\mathbf X(s)]$ and $\bm\sigma[s,\bm\be(s),\mathbf X(s)]$ are jointly measurable and continuous. For $s\in[0,T]$ the mapping $\bm\mu[s,\bm\be(s),\mathbf X(s)]:C^0([0,T],\mathbb{R}^{J},\mathbb{R}^{N\times J})\ra L(\mathbb{R}^n,\mathbb{R}^{J},\mathbb{R}^{N\times J})$ and $\bm\sigma[s,\bm\be(s),\mathbf X(s)]:\\C^0([0,T],\mathbb{R}^{J},\mathbb{R}^{N\times J})\ra L(\mathbb{R}^n,\mathbb{R}^{J},\mathbb{R}^{N\times J})$ are measurable with respect to the $\sigma$-algebra generated by the cylindrical sets with bases over the the time interval $[0,T]$ in continuous function vanishing at the infinity $C^0([0,T],\mathbb{R}^{J},\mathbb{R}^{N\times J})$, and the Borel $\sigma$-algebras in $\mathbb{R}^{J}$, $\mathbb{R}^{N\times J}$ and a linear functional $L(\mathbb{R}^n,\mathbb{R}^{J},\mathbb{R}^{N\times J})$ on a filtration $\mathcal F_s$ starting at time $s$, where time interval $[0,T]$ has been divided into $n$ small equal-lengthed subintervals. If above conditions hold, then for initial condition $\mathbf{X}_0\in\mathbb{R}^{(N\times J)\times 1}$ Krylov's theorem tells that, there exists a weak solution of coefficient dynamics represented by the Equation (\ref{0}) \cite{krylov2008}. The drift coefficient $\bm\mu[s,\bm\be(s),\mathbf{X}(s)]$ of the coefficient dynamics have different forms like for LASSO with $m$ covariates it is $\sum_{j'=1}^m|\be_{j'}(s)|$, ridge regression $\sum_{j'=1}^m\be_{j'}^2(s)$, standard $L^p$- norm $[\sum_{j'=1}^m|\be_{j'}(s)|^p]^{(1/p)}$, elastic net regression $(1-\a)||\bm\be(s)||_1+\a||\bm\be(s)||_2^2$ with $\a\in[0,1]$, group LASSO $\sum_{j'=1}^m \be_{j'}^T(s) K_{j'}(s)\be_{j'}(s)$ with $K_{j'}$ being a positive definite matrix, fused LASSO $\a\sum_{j'=1}^m |\be_{j'}(s)|+(1-\a)\sum_{j'=1}^m|\be_{j'}(s)-\be_{j'-1}(s)|$ and bridge regression $(\sum_{j'=1}^m\sqrt{|\be_{j'}(s)|})^2$ which we will discuss in examples. Furthermore, as we are concentrating in dynamic optimization, our objective is to
\begin{multline}\label{obj}
\min_{\{\be_j'\in \bm\be\}}\overline{\mathbf{X}}_O(s,\bm\be,\mathbf{X})= 
\min_{\{\be_j'\in \bm\be\}}\E \int_0^{T}
\sum_{i=1}^{N}\left[Y_i(s)-\sum_{j'=1}^J\beta_{j'}(s)X_{ij'}(s)\right]^2 ds,
\end{multline}
subject to the Equation (\ref{0}). To solve for the optimal coefficients we use Feynman-type path integral approach \cite{feynman1949} where we define a quantum Lagragian action function for small time interval $[s,\tau]\subseteq[0,T]$ as 
\begin{multline}\label{0.0}
\mathcal{L}_{s,\tau}(\mathbf{X})=\E_{s}\ \int_s^{\tau}\left\{\sum_{i=1}^{N}\left[Y_i(\nu)-\sum_{j'=1}^J\beta_{j'}(\nu)X_{ij'}(\nu)\right]^2d\nu\right. \\
\left.
\phantom{\int}
+\lambda[\Delta\bf U(\nu)-
\bm{\mu}[\nu,\bm\be(\nu),\mathbf{X}(\nu)]d\nu-
\bm{\sigma}[\nu,\bm\be(\nu),\mathbf{X}(\nu)]d\mathbf{B}(\nu)]
\right\},
\end{multline}
where $\lambda>0$ is the time independent penalization parameter. We will show the above integral in Equation (\ref{0.0}) measurable and then Feynman path integral of it is also  measurable in $\mathbb{R}^{N\times J}$ \cite{feynman1949}. Later part of this paper in Proposition \ref{p1} we will discuss about the closed form solutions of these coefficients under smoothing spline environment.

\section{Definitions and Assumptions}

\begin{defn}\label{d0}
Suppose a space $\mathcal{X}$ is Hausdorff. If for every point $x\in\mathcal{X}$ and every closed set $\mathcal Z\subseteq\mathcal X$ not containing $x$, there exists a continuous function $g_c:\mathcal X\ra[0,1]$ such that, $g_c(x)=1$ and $g_c(z)=0$ for all $z\in\mathcal Z$ then, $\mathcal{X}$ is completely regular \cite{bogachev}.
\end{defn}

\begin{defn}\label{d0.0}
For a family $\mathcal{M}$ of Radon measures on a topological space $\mathcal{X}$ if for every $\epsilon>0$, there exists a compact set $\kappa_\epsilon$ such that $|\rho|(\mathcal{X}\setminus\kappa_\epsilon)<\epsilon$ for all $\rho\in \mathcal{M}$ then $\mathcal X$ is called uniformly tight \cite{bogachev}.	
\end{defn}

Furthermore, from Definition \ref{d0.0} and Prohorov Theorem we know, if $\mathcal M$  is a family of Borel measures on $\mathcal{X}$ then every sequence $\{\rho_n\}_{n\geq 1}\subset\mathcal M$ contains a weakly convergent subsequence or $\mathcal M$ is uniformly tight and bounded \citep{bogachev, prokhorov1956}. In order to understand projective system of spaces let us assume $\mathcal{T}$ be a directed set and let $\{\mathbf X_n\}_{n\in\mathcal T}$ with $\gamma$ be a continuous mapping such that for two indices $n\geq m$ the condition $\gamma_{mn}:\mathbf{X}_n\ra\mathbf{X}_m $  and for $\eta\geq n\geq m$, $\gamma_{mn}\circ\gamma_{n\eta}=\gamma_{m\eta}$ hold. Furthermore, suppose $\mathbf X$ be a space such that mapping $\gamma_m:\mathbf X\ra\mathbf X_m$ is consistent with $\gamma_{nm}$ by the mapping $\gamma_m=\gamma_{mn}\circ\gamma_n$ for all $m\leq n$. Then $\mathbf X_m$ is the inverse limit space. As $\mathbf X=\mathbb{R}^\infty$ is an example of this space, the dimension of our independent variables $\mathbf X_{nJ}=\mathbb{R}^{n\times J}$ consists of all sequences of the form $(X_1,...,X_{nJ},0,...,0)$, and $\gamma_{nJk}$ and $\gamma_{nJ}$ are natural projections. Now consider spaces $\mathbf X_n$ are equipped with Borel $\sigma$-algebra $\mathcal B_n$  and measures $\rho_n$ on $\mathcal B_n$ such that $\gamma_{mn}$ are measurable. Then for $m\leq n$
\begin{equation*}
\gamma_{mn}(\rho_n):=\rho_n\circ \gamma_{mn}^{-1}=\rho_m,
\end{equation*}
is a necessary condition. Furthermore, for $\rho$ is a Radon measure on $\mathbf X$, $\rho\circ\gamma_n^{-1}=\rho_n,\ \forall n$ exists iff for any $\epsilon>0$, $\exists \kappa_\epsilon\subset\mathbf X$ with $\rho_n(\gamma_n(\kappa_\epsilon)) \geq 1-\epsilon, \ \forall n$ \cite{bogachev}. We use this result to prove Lemma \ref{l0.0}.

\begin{as}\label{as0.0}
For time interval $[s,s+\epsilon]\subset[0,T]$, where $\epsilon\downarrow 0$ the filtration space starting at time $s$ denoted by $\mathcal{F}_s$, is a vector lattice of on the non-empty set $\Omega$ such that for point-transition $\Psi_s(\mathbf{X})$ and quantum Lagrangian $\mathcal{L}_{s,s+\epsilon}$,
\begin{equation*}
\Psi_{s,s+\epsilon}(\mathbf{X})=
\frac{1}{N_s^{\tilde f}}\int_{\mathbb{R}^{N\times J}}\tilde fd\mathbf{X},
\end{equation*}
where $\tilde f=\exp
\left[-\epsilon \mathcal{L}_{s,s+\epsilon}(\mathbf{X})\right] \Psi_s(\mathbf{X})$ and $N_s^{\tilde f}>0$ is a normalizing constant of $\tilde f\in\mathcal{F}_s$. For another function $\tilde g\in\mathcal{F}_s$ with normalizing constant $N_s^{\tilde g}>0$ define
\begin{equation*}
\tilde{\Psi}_{s,s+\epsilon}(\mathbf{X})=
\frac{1}{N_s^{\tilde g}}\int_{\mathbb{R}^{N\times J}}\tilde gd\mathbf{X},
\end{equation*}
such that $\max(\tilde f,\tilde g)\in\mathcal{F}_s$, $\min(\tilde f,\tilde g)=-\max(\tilde f,\tilde g)$ and $|\tilde f|\in\mathcal{F}_s$.
\end{as}

\begin{as}\label{as0.1}
The set of all bounded functions $\mathcal F_s^+$ of $\tilde f$ such that for a non-negative increasing sequence $\tilde f_k\in\mathcal{F}_s$ the condition $\tilde f=\lim_{k\ra\infty}\tilde f_k$ holds. As the sequence $\{\tilde f_k\}$ is uniformly bounded, we assume the sequence $\{\Psi_{s,s+\epsilon}^k\}$	is increasing and bounded where,
\begin{equation*}
\Psi_{s,s+\epsilon}^k(\mathbf{X})=
\frac{1}{N_s^{\tilde f_k}}\int_{\mathbb{R}^{N\times J}}\tilde f_kd\mathbf{X}.
\end{equation*}
Assume $\Psi_{s,s+\epsilon}(\mathbf X)=\lim_{k\ra\infty}\Psi_{s,s+\epsilon}^k(\mathbf{X})$. Then For all $\tilde f,\tilde g\in\mathcal{F}_s^+$ and $\tilde f\leq\tilde g$ there exists a measure $(N_s^{\tilde f})^{-1}d\mathbf X$ such that following conditions hold,\\
	$1$.  $\Psi_{s,s+\epsilon}(\mathbf X)\leq\tilde{\Psi}_{s,s+\epsilon}(\mathbf X)$;\\
	$2$. $\Psi_{s,s+\epsilon}^*(\mathbf X)=\Psi_{s,s+\epsilon}(\mathbf X)+\tilde{\Psi}_{s,s+\epsilon}(\mathbf X)$, where
	\begin{equation*}
	\Psi_{s,s+\epsilon}^*(\mathbf{X})=
	\frac{1}{N_s^{(\tilde f+\tilde g)}}\int_{\mathbb{R}^{N\times J}}(\tilde f+\tilde g)d\mathbf{X}.
	\end{equation*}\\
	$3$. For a constant $c\in[0,\infty)$, $\Psi_{s,s+\epsilon}^c(\mathbf X)=c\Psi_{s,s+\epsilon}(\mathbf X)$ where,
	\begin{equation*}
	\Psi_{s,s+\epsilon}^c(\mathbf{X})=
	\frac{1}{N_s^{(c\tilde f)}}\int_{\mathbb{R}^{N\times J}}(c\tilde f)d\mathbf{X}.
	\end{equation*}\\
	$4$. For all $\min(\tilde f,\tilde g)\in\mathcal{F}_s^+$ and $\max(\tilde f,\tilde g)\in\mathcal{F}_s^+$ we have $\Psi_{s,s+\epsilon}(\mathbf X)+\tilde{\Psi}_{s,s+\epsilon}=\Psi_{s,s+\epsilon}^{\min}(\mathbf X)+\Psi_{s,s+\epsilon}^{\max}(\mathbf X)$ where
	\begin{equation*}
	\Psi_{s,s+\epsilon}^{\min}(\mathbf{X})=
	\frac{1}{N_s^{\min(\tilde f,\tilde g)}}\int_{\mathbb{R}^{N\times J}}\min(\tilde f,\tilde g)d\mathbf{X}
	\end{equation*}
	and
	\begin{equation*}
	\Psi_{s,s+\epsilon}^{\max}(\mathbf{X})=
	\frac{1}{N_s^{\max(\tilde f,\tilde g)}}\int_{\mathbb{R}^{N\times J}}\max(\tilde f,\tilde g)d\mathbf{X}.
	\end{equation*}\\
	$5$. $\lim_{k\ra\infty}\tilde f_k\in\mathcal{F}_s^+$ for every uniformly bounded sequence of $\tilde f_k\in\mathcal{F}_s^+$, and one has $\Psi_{s,s+\epsilon}^{\lim}(\mathbf X)=\lim_{k\ra \infty}\Psi_{s,s+\epsilon}^k(\mathbf X)$ where
	\begin{equation*}
	\Psi_{s,s+\epsilon}^{\lim}(\mathbf{X})=
	\frac{1}{N_s^{\lim_{k\ra\infty}\tilde f_k}}\int_{\mathbb{R}^{N\times J}}\lim_{k\ra\infty}\tilde f_kd\mathbf{X}.
	\end{equation*}
\end{as}

\begin{as}\label{as0}
	For $T>0$, let ${\bm{\mu}}(s,\bm\be,\mathbf{X}):C^0([0,T],\mathbb{R}^{J},\mathbb{R}^{N\times J})\ra L(\mathbb{R}^n,\mathbb{R}^{J},\mathbb{R}^{N\times J})$ and $\bm{\sigma}(s,\bm\be,\mathbf{X}):C^0([0,T],\mathbb{R}^{J},\mathbb{R}^{N\times J})\ra L(\mathbb{R}^n,\mathbb{R}^{J},\mathbb{R}^{N\times J})$ be some measurable function and, for some positive constant $K_1$ and, $\mathbf{X}\in\mathbb{R}^{N\times J}$ we have linear growth of $\bm\be$ as
	\[
	|{\bm{\mu}}(s,\bm\be,\mathbf{X})|+
	|\bm{\sigma}(s,\bm\be,\mathbf{X})|\leq 
	K_1(1+|\mathbf{X}|),
	\]
	such that, there exists another positive, finite, constant $K_2$ and for a different vector 
	$\widetilde{\mathbf{X}}_{(N\times J)\times 1}$ such that the Lipschitz condition,
	\[
	|{\bm{\mu}}(s,\bm\be,\mathbf{X})-
	{\bm{\mu}}(s,\bm\be,\widetilde{\mathbf{X}})|+|\bm{\sigma}(s,\bm\be,\mathbf{X})-\bm{\sigma}(s,\bm\be,\widetilde{\mathbf{X}})|
	\leq K_2\ |\mathbf{X}-\widetilde{\mathbf{X}}|,\notag
	\]
	$ \widetilde{\mathbf{X}}\in\mathbb{R}^{N\times J}$ is satisfied and
	\[
	|{\bm{\mu}}(s,\bm\be,\mathbf{X})|^2+
	\|\bm{\sigma}(s,\bm\be,\mathbf{X})\|^2\leq K_2^2
	(1+|\widetilde{\mathbf{X}}|^2),
	\]
	where 
	$\|\bm{\sigma}(s,\bm\be,\mathbf{X})\|^2=
	\sum_{i=1}^N \sum_{j=1}^N|{\sigma^{ij}}(s,\bm\be,\mathbf{X})|^2$.
\end{as}

\begin{as}\label{as1}
	There exists a probability space $(\Omega,\mathcal{F}_s^{\mathbf X},\mathcal{P})$ with sample space $\Omega$, filtration at time $s$ of independent variable ${\mathbf{X}}$ as $\{\mathcal{F}_s^{\mathbf{X}}\}\subset\mathcal{F}_s$, a probability measure $\mathcal{P}$ and a $p$-dimensional $\{\mathcal{F}_s\}$ Brownian motion $\mathbf{B}$ where the measure of the regression coefficient $\bm\be$ is an $\{\mathcal{F}_s^{\mathbf{X}}\}$ adapted process such that Assumption \ref{as0} holds.
\end{as}

\section{Main Results}
The objective function is,
\begin{multline}\label{2}
\min_{\{\be_j'\in \bm\be\}}\overline{\mathbf{X}}_O(s,\bm\be,\mathbf{X})= \\
\min_{\{\be_j'\in \bm\be\}}\E \int_0^{T}
\sum_{i=1}^{N}\left[Y_i(s)-\sum_{j'=1}^J\beta_{j'}(s)X_{ij'}(s)\right]^2 ds,
\end{multline}
In Equation (\ref{2}), $\beta_j$ is the coefficient of independent variable $X_{ij}$ for all $i=1,...,N$ and $j'=1,...,J$.

\begin{lem}\label{l0.0}
Suppose time interval $[0,T]$ and $\mathbf{R}^{N\times J}$ are completely regular space such that the space $\mathcal{T}=[0,T]\times \mathbb{R}^{N\times J}$ is also completely regular and all the compact subsets in it have Euclidean metrics	and let a measure $\rho_n\in\mathcal{M}(\Omega\times \mathcal{T})$ converges towards a measure $\rho\in\mathcal{M}(\Omega\times\mathcal{T})$ and is uniformly bounded in the variation norm. If the projections of the measure $|\rho_n|$ and $|\rho|$ on $\mathcal{T}$ are uniformly tight and the projections of the measures $|\rho_n|$ on $\Omega$ are uniformly countably additive, then
\begin{equation}\label{0.1}
\lim_{n\ra\infty}\E_s\int_{s}^\tau \hat f d\rho_n=\E\int_{0}^T \hat f d\rho,
\end{equation}
where $n$ is the total number of small equal in length subintervals $[s,\tau]$ of $[0,T]$ and the continuous bounded $\mathcal{P}\otimes\mathcal{B}(\mathcal{T})$-Borel measurable function $\hat f$ such that,
\begin{multline*}
\int_{s}^\tau\hat f d\rho_n=\int_s^{\tau}\left\{\sum_{i=1}^{N}\left[Y_i(\nu)-\sum_{j'=1}^J\beta_{j'}(\nu)X_{ij'}(\nu)\right]^2d\nu\right. \\
\left.
\phantom{\int}
+\lambda[\Delta\bf U(\nu)-
\bm{\mu}[\nu,\bm\be(\nu),\mathbf{X}(\nu)]d\nu-
\bm{\sigma}[\nu,\bm\be(\nu),\mathbf{X}(\nu)]d\mathbf{B}(\nu)]
\right\},
\end{multline*}
where $\mathcal{P}$ is the probability measure on the Borel $\sigma$-algebra $\mathcal{B}(\mathcal{T})$.
\end{lem}

\begin{lem}\label{l0.1}
Suppose, for $\epsilon\downarrow 0$, $\Psi_{s,s+\epsilon}$ approximated to a linear function on $\mathcal{F}_s$ within the small time interval $[s,s+\epsilon]$ such that, Assumptions \ref{as0.0}-\ref{as1}, Lemma \ref{l0.0} hold and for $\tilde f\geq 0$ we have $\Psi_{s,s+\epsilon}(\mathbf X)$ , $\lim_{k\ra\infty}\Psi_{s,s+\epsilon}^k(\mathbf X)\ra 0$ for every monotonically decreasing sequence $\tilde f_k\in\mathcal{F}_s$. Then there exists a unique measure $N_s^{-1}d\mathbf{X}$ generated by the filtration $\mathcal{F}_s^{\mathbf X}$ starting at $\mathbf X_0\in\mathbb{R}^{N\times J}$ such that $\mathcal F_s^{\mathbf X}\subseteq\mathcal F_s$ and
\begin{equation*}
\Psi_{s,s+\epsilon}(\mathbf X)=\frac{1}{N_s}\int_{\mathbb R^{N\times J}}\tilde f d\mathbf X,\ \forall \tilde f\in \mathcal F_s,
\end{equation*}
where $\tilde f=\exp
\left[-\epsilon \mathcal{L}_{s,s+\epsilon}(\mathbf{X})\right]
\Psi_s(\mathbf{X})$.
\end{lem}

\begin{prop}\label{p0}
If the objective is to minimize Equation (\ref{2}) subject to the error dynamics 
\begin{equation}\label{run}
d\mathbf{\bf U}(s)=\bm{\mu}[s,\bm\be(s),\mathbf{X}(s)]ds+
\bm{\sigma}[s,\bm\be(s),\mathbf{X}(s)]d\mathbf{B}(s),
\end{equation} 
with Assumptions \ref{as0.0}-\ref{as1} and, Lemmas \ref{l0.0}, \ref{l0.1}, then under continuous time, for $\{i,j\}=\{1,...,N\}^2,$ $j'=1,...J$, $X_{ij'}$'s regression coefficient  is found by solving the Equation 
\begin{multline*}
2\sum_{i=1}^{N}\left[Y_i(s)-\sum_{j'=1}^J\beta_{j'}(s)X_{ij'}(s)\right] X_{ij'}(s)\\
-\frac{\partial g[s,\mathbf{X}(s)]}{\partial \mathbf{X}}
\frac{\partial \bm{\mu}[s,\bm\be(s),\mathbf{X}(s)]}{\partial \bm\be(s)}	\frac{\partial \bm\be(s)}{\partial \be_{j'}(s) } \\ 
-\mbox{$\frac{1}{2}$}
\sum_{i=1}^N\sum_{j=1}^N
\frac{\partial \bm\sigma^{ij}[s,\bm\be(s),\mathbf{X}(s)]}
{\partial \bm\be(s)}
\frac{\partial \bm\be(s)}{\partial \be_{j'} }
\frac{\partial^2 g[s,\mathbf{X}(s)]}{\partial{X_{ij'}\partial X_{jj'}}}=0,
\end{multline*}
for $\beta_{j'}$, with initial condition  $\mathbf{X}_{0_{(N\times J)\times 1}}$,
where $g[s,\mathbf{X}(s)]\in C^2\left([0,T]\times \mathbb{R}^{N\times J}\right)$ with $\mathbf{I}(s)=g[s,\mathbf{X}(s)]$ is a positive, non-decreasing penalization function vanishing at infinity which substitutes the coefficient dynamics such that, $\mathbf{I}(s)$ is an It\^o process. 
\end{prop}

\begin{example}
(LASSO). Consider the dynamic objective function expressed in the Equation (\ref{obj}) subject to the error dynamics
\begin{equation*}
d{\bf U}(s)=\sum_{j'=1}^m|\be_{j'}(s)|ds+2\sum_{i=1}^N\sum_{j'=1}^m\be_{j'}(s)X_{ij'}(s) dB(s),
\end{equation*}
where $B(s)$ is the constant Brownian motion of this system. The main reason of not taking a squared root in the diffusion coefficient is $\be_{j'}(s)$ is small in magnitude. We further assume independent variables evolves exponentially. Therefore, for a positive penalization parameter $\lambda^*$, we assume $g(s, X_{ij'})=\lambda^*\exp(s X_{ij'})$ where $\frac{\partial}{\partial X_{ij'}}g(s,X_{ij'})=sg(s, X_{ij'})$ and $\frac{\partial^2}{\partial X_{ij'}^2}g(s, X_{ij'})=\\s^2g(s,X_{ij'})$. Furthermore, without loss of generality we assume $m=J$ and our main concern is to find the optimal coefficient, we assume $\be_k\neq0$ for any $k=1,...,J$. Therefore, $\frac{\partial}{\partial \be_k} |\be_k|=\frac{\be_k}{|\be_k|}$ which is $-1$ for all $\be_k<0$ and $1$ for all $\be_k>0$. By using Proposition \ref{p0} we have,
\begin{multline*}
2\sum_{i=1}^{N}\left[Y_i(s)-\be_k(s) X_{ik}(s)-\sum_{j'=1}^{J-1}\beta_{j'}(s)X_{ij'}(s)\right] X_{ij'}(s)\\
-sg(s,X_{ij'})\frac{\be_k}{|\be_k|}-s^2\sum_{i=1}^N X_{ij'}(s)g(s,X_{ij'})=0,
\end{multline*}
which yields,
\begin{multline*}
\be_k=\frac{1}{2\sum_{i=1}^N X_{ik}^2(s)}\bigg\{2\sum_{i=1}^N X_{ij'}(s)Y_i(s)\\
-2\sum_{i=1}^N\sum_{j'=1}^{J-1}\be_{j'}(s)X_{ij'}(s)-s\left[g(s,X_{ij'})+s\sum_{i=1}^N X_{ij'}g(s,X_{ij'})\right]\bigg\},
\end{multline*}
for all $\be_k>0$ and $\sum_{i=1}^N X_{ik}^2(s)\neq 0$ and,
\begin{multline*}
\be_k=\frac{1}{2\sum_{i=1}^N X_{ik}^2(s)}\bigg\{2\sum_{i=1}^N X_{ij'}(s)Y_i(s)\\
-2\sum_{i=1}^N\sum_{j'=1}^{J-1}\be_{j'}(s)X_{ij'}(s)+s\left[g(s,X_{ij'})-s\sum_{i=1}^N X_{ij'}g(s,X_{ij'})\right]\bigg\},
\end{multline*}
for all $\be_k<0$.
\end{example}

\begin{example}
(Ridge regression). Consider again objective function in Equation (\ref{obj}) subject to
\begin{equation*}
d{\bf U}(s)=\sum_{j'=1}^J\be_{j'}^2(s)ds+2\sum_{i=1}^N\sum_{j'=1}^J\be_{j'}(s)X_{ij'}(s) dB(s).
\end{equation*}
Assuming $g(s, X_{ij'})=\lambda^*\exp(s X_{ij'})$ for all $\sum_{i=1}^N X_{ik}^2+sg(s,X_{ik})\neq 0$, where $k=1,...,J$, Proposition \ref{p0} determines the regression coefficient under ridge regression as 
\begin{equation*}
\be_k=\frac{2\left[\sum_{i=1}^NX_{ij'}(s)Y_i(s)-\sum_{j'=1}^{J-1}\be_{j'} X_{ij'}^2\right]-s^2\sum_{i=1}^N X_{ij'}(s)g(s,X_{ij'})}{2\left[\sum_{i=1}^N X_{ik}^2(s)+sg(s,X_{ik})\right]}.
\end{equation*}
\end{example}

\begin{example}
(Standard $L^p$-norm). In this framework for all $p\neq 0$ let us assume the error dynamics as 
\begin{equation*}
d{\bf U}(s)=\left[\sum_{j'=1}^J|\be_{j'}(s)|^p\right]^{\frac{1}{p}}ds+2\sum_{i=1}^N\sum_{j'=1}^J\be_{j'}(s) X_{ij'}(s)dB(s),
\end{equation*}
where $dB(s)$ is the constant Brownian motion in this system such that $\be_{j'}\neq 0$ for all $j'=1,...,J$. If we minimize the Equation (\ref{obj}) subject to the above coefficient dynamics, Proposition \ref{p0} with $g(s, X_{ij'})=\lambda^*\exp(s X_{ij'})$ gives
\begin{multline*}
2\sum_{i=1}^N Y_i(s)X_{ij'}(s)-2\sum_{i=1}^N\sum_{j'=1}^J\be_{j'}(s) X_{ij'}(s)\\
-sg(s,X_{ij'})\left[\sum_{j'=1}^J|\be_{j'}(s)|^p\right]^{\frac{1}{p}-1}\sum_{j'=1}^J\be_{j'}|\be_{j'}|^{p-2}-s^2\sum_{i=1}^N X_{ij'}g(s,X_{ij'})=0.
\end{multline*}
Hence, for $k=1,...,J$ we have,
\begin{multline*}
2\sum_{i=1}^N Y_i(s)X_{ij'}(s)-2\be_k(s)\sum_{i=1}^NX_{ik}^2(s)\\-2\sum_{i=1}^N\sum_{j'=1}^{J-1}\be_{j'}(s) X_{ij'}(s)-sg(s,X_{ik})\left[|\be_{k}(s)|^p\right]^{\frac{1}{p}-1}\be_{k}(s)|\be_{k}(s)|^{p-2}\\
-sg(s,X_{ij'})\left[\sum_{j'=1}^{J-1}|\be_{j'}(s)|^p\right]^{\frac{1}{p}-1}\sum_{j'=1}^{J-1}\be_{j'}(s)|\be_{j'}(s)|^{p-2}-s^2\sum_{i=1}^N X_{ij'}g(s,X_{ij'})=0.
\end{multline*}
Furthermore, For all $\be_k>0$, $\be_{j'}>0$  and $\sum_{i=1}^N X_{ik}(s)\neq 0$ we have,
\begin{align*}
\be_k& = \frac{1}{2\sum_{i=1}^NX_{ik}^2(s)}\biggr\{2\sum_{i=1}^NY_i(s)X_{ij'}(s)-2\sum_{i=1}^N\sum_{j'=1}^{J-1}\be_{j'}(s)X_{ij'}^2(s)\\ &-sg(s,X_{ij'})\left[\sum_{j'=1}^J\be_{j'}^p(s)\right]^{\frac{1}{p}-1}\sum_{j'=1}^{J-1}\be_{j'}^{p-1}(s)-sg(s,X_{ik})-s^2\sum_{i=1}^NX_{ij'}(s)g(s,X_{ij'})\biggr\},
\end{align*}
and when $\be_k<0$ and $\be_{j'}<0$ for all $j',k=1,...,J$, then
\begin{multline*}
\be_k = \frac{1}{2\sum_{i=1}^NX_{ik}^2(s)}\biggr\{s^2\sum_{i=1}^NX_{ij'}(s)g(s,X_{ij'})\\-2\sum_{i=1}^NY_i(s)X_{ij'}(s) -2\sum_{i=1}^N\sum_{j'=1}^{J-1}\be_{j'}(s)X_{ij'}^2(s)\\-sg(s,X_{ij'})\left[\sum_{j'=1}^J\be_{j'}^p(s)\right]^{\frac{1}{p}-1}\sum_{j'=1}^{J-1}\be_{j'}^{p-1}(s)-sg(s,X_{ik})\biggr\}.
\end{multline*}
\end{example}

\begin{example}
(Elastic net regression). In this framework for all $\a\in[0,1]$ and $\be_{j'}\neq 0,\forall j'=1,...,J$ suppose the error dynamics is 
\begin{equation*}
d{\bf U}(s)=\left\{(1-\a)\sum_{j'=1}^J|\be_{j'}(s)|+\a\sum_{j'=1}^J\be_{j'}^2(s)\right\}ds+2\sum_{i=1}^N\sum_{j'=1}^J\be_{j'}(s) X_{ij'}(s)dB(s),
\end{equation*}
where $dB(s)$ is the constant Brownian motion in this system. If we minimize the Equation (\ref{obj}) subject to the above coefficient dynamics, Proposition \ref{p0} with $g(s, X_{ij'})=\lambda^*\exp(s X_{ij'})$ gives
\begin{multline*}
2\sum_{i=1}^NY_i(s)X_{ij'}(s)-2\be_k(s)\sum_{i=1}^N X_{ik}^2(s)-2\sum_{i=1}^N\sum_{j'=1}^{J-1}\be_{j'}(s)X_{ij'}^2(s)\\
-2sg(s,X_{ij'})\left[(1-\a)\frac{\be_k(s)}{|\be_k(s)|}+2\a\be_k(s)\right]-s^2\sum_{i=1}^N X_{ij'}(s) g(s,X_{ij'})=0,
\end{multline*}
for $\be_k>0$, $\be_{j'}>0$ and $\sum_{i=1}^N X_{ik}^2-s\a g(s,X_{ij'})\neq 0$ which gives us
\begin{multline*}
\be_k=\frac{1}{2\left[\sum_{i=1}^N X_{ik}^2(s)-s\a g(s,X_{ij'})\right]}\biggr[2\sum_{i=1}^NY_i(s)X_{ij'}(s)\\-2\sum_{i=1}^N\sum_{j'=1}^{J-1}\be_{j'}(s)X_{ij'}^2(s)-s(1-\a)g(s,X_{ij'})-s^2\sum_{i=1}^N X_{ij'}(s)g(s,X_{ij'})\biggr],
\end{multline*}
and, for $\be_k<0$, $\be_{j'}<0$
\begin{multline*}
\be_k=\frac{1}{2\left[\sum_{i=1}^N X_{ik}^2(s)+s\a g(s,X_{ij'})\right]}\biggr[s^2\sum_{i=1}^N X_{ij'}(s)g(s,X_{ij'})\\-2\sum_{i=1}^NY_i(s)X_{ij'}(s)-2\sum_{i=1}^N\sum_{j'=1}^{J-1}\be_{j'}(s)X_{ij'}^2(s)-s(1-\a)g(s,X_{ij'})\biggr].
\end{multline*}
\end{example}

\begin{example}
(Fused LASSO). In this framework for all $\a\in(0,1)$ let us assume the coefficient dynamics as 
\begin{multline*}
d{\bf U(s)}=\left[\a\sum_{j'=1}^J|\be_{j'}(s)|+(1-\a)\sum_{j'=0}^J|\be_{j'}(s)-\be_{j'-1}(s)|\right]ds\\+2\sum_{i=1}^N\sum_{j'=1}^J\be_{j'}(s) X_{ij'}(s)dB(s),
\end{multline*}
where $dB(s)$ is the constant Brownian motion in this system such that $\be_{j'}\neq 0$ for all $j'=1,...,J$. For a function $g(s,X_{ij'})=\lambda^*\exp(sX_{ij'})$ Proposition \ref{p0} yields,
\begin{multline*}
2\sum_{i=1}^NY_i(s)X_{ij'}(s)-2\be_k(s)\sum_{i=1}^N X_{ik}^2(s)-2\sum_{j'=1}^{J-1}\be_{j'}(s)X_{ij'}^2(s)\\
-sg(s,X_{ik})\left[\a\frac{\be_k(s)}{|\be_k(s)|}+(1-\a)\frac{\be_k(s)-\be_{k-1}(s)}{|\be_k(s)-\be_{k-1}(s)|}\right]-s^2\sum_{i=1}^N X_{ij'}(s)g(s,X_{ij'})=0.
\end{multline*}
Furthermore, for $\sum_{i=1}^N X_{ik}^2(s)\neq 0$ if $\be_k>0,\ \be_{j'}>0$ for all $k=1,...,J$  such that $\be_k>\be_{k-1}$ then
\begin{multline*}
\be_k=\frac{1}{2\sum_{i=1}^N X_{ik}^2(s)}\biggr[2\sum_{i=1}^N Y_i(s) X_{ij'}(s)-2\sum_{j'=1}^{J-1}\be_{j'}(s)X_{ij'}^2(s)\\-sg(s,X_{ik})-s^2\sum_{i=1}^N X_{ij'}(s)g(s,X_{ij'})\biggr],
\end{multline*}
if $\be_k>0,\ \be_{j'}>0$ such that $\be_k<\be_{k-1}$ then for $\sum_{i=1}^N X_{ik}^2\neq 0$ we have,
\begin{multline*}
\be_k=\frac{1}{2\sum_{i=1}^N X_{ik}^2(s)}\biggr[2\sum_{i=1}^N Y_i(s) X_{ij'}(s)-2\sum_{j'=1}^{J-1}\be_{j'}(s)X_{ij'}^2(s)\\+s(1-2\a)g(s,X_{ik})-s^2\sum_{i=1}^N X_{ij'}(s)g(s,X_{ij'})\biggr],
\end{multline*}
and finally, if $\be_k<0,\ \be_{j'}<0$ such that $\be_k\neq\be_{k-1}$ then
\begin{multline*}
\be_k=\frac{1}{2\sum_{i=1}^N X_{ik}^2(s)}\biggr[s^2\sum_{i=1}^N X_{ij'}(s)g(s,X_{ij'})-2\sum_{i=1}^N Y_i(s) X_{ij'}(s)\\-2\sum_{j'=1}^{J-1}\be_{j'}(s)X_{ij'}^2(s)-sg(s,X_{ik})\biggr].
\end{multline*} 
\end{example}

\begin{example}
(Bridge regression). For all  $\be_{j'}\neq 0,\forall j'=0,...,J$ suppose the error dynamics is 
\begin{equation*}
d{\bf U}(s)=\left\{\sum_{j'=1}^J|\be_{j'}(s)|^{\frac{1}{2}}\right\}^2ds+2\sum_{i=1}^N\sum_{j'=1}^J\be_{j'}(s) X_{ij'}(s)dB(s),
\end{equation*}
where $dB(s)$ is the constant Brownian motion in this system. If we minimize the Equation (\ref{obj}) subject to the above coefficient dynamics, Proposition \ref{p0} with $g(s, X_{ij'})=\lambda^*\exp(s X_{ij'})$ gives
\begin{multline*}
2\sum_{i=1}^N Y_i(s)X_{ij'}(s)-2\sum_{i=1}^N\sum_{j'=1}^{J}\be_{j'}(s)X_{ij'}^2(s)\\-s\be_{j'}(s)|\be_{j'}(s)|^{-\frac{3}{2}}g(s,X_{ij'})\sum_{j'=1}^{J}|\be_{j'}(s)|^{\frac{1}{2}}-s^2\sum_{i=1}^N X_{ij'}(s) g(s,X_{ij'})=0.
\end{multline*}
Furthermore, for all $k=1,...,J$, $\be_k>0$, $\be_{j'}>0$ and $\sum_{i=1}^N X_{ik}^2\neq 0$ we have,
\begin{multline*}
\be_k=\frac{1}{2\sum_{i=1}^N X_{ik}^2(s)}\biggr[2\sum_{i=1}^NY_i(s)X_{ij'}(s)-2\sum_{i=1}^N\sum_{j'=1}^{J-1}\be_{j'}(s)X_{ij'}^2(s)\\
-sg(s,X_{ik})-s\be_{j'}^{-\frac{1}{2}}(s)g(s,X_{ij'})\sum_{j'=1}^{J-1}\be_{j'}^{\frac{1}{2}}(s)-s^2\sum_{i=1}^NX_{ij'}(s)g(s,X_{ij'})\biggr],
\end{multline*}
and for $\be_k<0$, $\be_{j'}<0$ and $\sum_{i=1}^N X_{ik}^2\neq 0$ we have,
\begin{multline*}
\be_k=\frac{1}{2\sum_{i=1}^N X_{ik}^2(s)}\biggr[s^2\sum_{i=1}^NX_{ij'}(s)g(s,X_{ij'})-2\sum_{i=1}^NY_i(s)X_{ij'}(s)\\-2\sum_{i=1}^N\sum_{j'=1}^{J-1}\be_{j'}(s)X_{ij'}^2(s)-sg(s,X_{ik})-s\be_{j'}^{-\frac{1}{2}}(s)g(s,X_{ij'})\sum_{j'=1}^{J-1}\be_{j'}^{\frac{1}{2}}(s)\biggr].
\end{multline*}
\end{example}

\begin{example}
(Group LASSO). For an $m$-dimensional coefficient vector $\bm\be_{j'}$ with $\mathbf K_{j'}$, an ${m\times m}$-dimensional positive definite matrix  assume the error  dynamics is,
\begin{equation*}
d{\bf U}(s)=\left[\sum_{j'=1}^J\bm\be_{j'}^T(s)\mathbf K_j(s)\bm\be_{j'}(s)\right]ds+2\sum_{i=1}^N\sum_{j'=1}^J\bm\be_{j'}^T\mathbf X_{ij'}(s)d\mathbf B(s),
\end{equation*}
where $\be_{j'}^T$ is the transposition of $\be_{j'}$,  $\mathbf X_{ij'}$ an $m\times m$-dimensional matrix and $d\mathbf B(s)$ is an $m$-dimensional Brownian motion. Using an $m$-dimensional vector valued function $g(s,\mathbf X_{ij'})=\lambda^*\exp(s,\mathbf X_{ij'})$ and Proposition \ref{p0} we get coefficient vector as,
\begin{multline*}
\bm\be_k=\mbox{$\frac{1}{2}$}\left\{\sum_{i=1}^N\mathbf X_{ik}^T(s)\mathbf X_{ik}(s)+\left[\mathbf K_k(s)+\mathbf K_k^T(s)\right]g(s,\mathbf X_{ik})\right\}^{-1}\times\\
\biggr[2\sum_{i=1}^N\mathbf Y_i(s)\mathbf X_{ij'}(s)-2\sum_{i=1}^N\sum_{j'=1}^{J-1}\bm\be_{j'}(s)\mathbf X_{ij'}^T(s)\mathbf X_{ij'}(s)\\
-sg(s,\mathbf X_{ij'})\sum_{j'=1}^{J-1}\left[\mathbf K_{j'}(s)+\mathbf K_{j'}^T(s)\right]\bm\be_{j'}(s)-s^2\sum_{i=1}^N\mathbf X_{ij'}(s)g(s,\mathbf X_{ij'})\biggr],
\end{multline*}
Such that $\left[\sum_{i=1}^N\mathbf X_{ik}^T(s)\mathbf X_{ik}(s)+\left[\mathbf K_k(s)+\mathbf K_k^T(s)\right]g(s,\mathbf X_{ik})\right]^{-1}$ exists and $\bm\be_k$ is an $m$-dimensional vector where $k=1,...,J$.
\end{example}

\begin{prop}\label{p1}
Suppose, under the system of smoothing spline regression our objective is to,
\begin{multline}\label{2.1}
\min_{\{\be_j'\in \bm\be\}}\overline{\mathbf{X}}_S(s,\bm\be,\mathbf{X})= \\
\min_{\{\be_j'\in \bm\be\}}\E \int_0^{T}
\sum_{i=1}^{N}\left[Y_i(s)-\sum_{j'=1}^J\beta_{j'}(s)h[X_{ij'}(s)]\right]^2 ds,
\end{multline}
subject to the error dynamics represented by the Equation (\ref{0}), where $h$ is a dynamic $C^2$-basis function such that Assumptions \ref{as0.0}- \ref{as1}, Lemmas \ref{l0.0} and \ref{l0.1} hold. Then under continuous time, for $\{i,j\}=\{1,...,N\}^2,$ $j'=1,...J$, $h[X_{ij'}]$'s regression coefficient  is found by solving the Equation 
\begin{multline*}
2\sum_{i=1}^{N}\left[Y_i(s)-\sum_{j'=1}^J\beta_{j'}(s)h[X_{ij'}(s)]\right] h[X_{ij'}(s)]\\
-\frac{\partial g^*[s,\mathbf{X}(s)]}{\partial \mathbf{X}}
\frac{\partial \bm{\mu}[s,\bm\be(s),\mathbf{X}(s)]}{\partial \bm\be(s)}	\frac{\partial \bm\be(s)}{\partial \be_{j'}(s) } \\ 
-\mbox{$\frac{1}{2}$}
\sum_{i=1}^N\sum_{j=1}^N
\frac{\partial \bm\sigma^{ij}[s,\bm\be(s),\mathbf{X}(s)]}
{\partial \bm\be(s)}
\frac{\partial \bm\be(s)}{\partial \be_{j'} }
\frac{\partial^2 g[s,\mathbf{X}(s)]}{\partial{X_{ij'}\partial X_{jj'}}}=0,
\end{multline*}
for $\beta_{j'}$, with initial condition  $\mathbf{X}_{0_{(N\times J)\times 1}}$,
where $g^*[s,\mathbf{X}(s)]\in C^2\left([0,T]\times \mathbb{R}^{N\times J}\right)$ with $\mathbf{I}^*(s)=g^*[s,\mathbf{X}(s)]$ is a positive, non-decreasing penalization function vanishing at infinity which substitutes the coefficient dynamics such that, $\mathbf{I}^*(s)$ is an It\^o process. 
\end{prop}

\begin{example}\label{e8}
(Cubic smoothing spline) Consider the objective function in Equation (\ref{2.1}) where $h[X_{ij'}(s)]=X_{ij'}(s)+X_{ij'}^2(s)+X_{ij'}^3(s)$ subject to the error dynamics
\begin{equation*}
d{\bf U}(s)=\left[2\be_{j'}(s)+6\be_{j'}(s)X_{ij'}(s)\right]ds+2\sum_{i=1}^N\be_{j'}(s)X_{ij'}(s)dB(s),
\end{equation*}
where $B(s)$ is the Brownian motion under cubic smoothing spline. For a penalization parameter $\lambda^*$ if we assume $g^*(s,X_{ij'})=\lambda^*\exp(sX_{ij'})$ then Proposition \ref{p1} gives us,
\begin{multline*}
0=2\sum_{i=1}^N\left\{Y_i(s)-\sum_{j'=1}^J\be_{j'}(s)\left[X_{ij'}(s)+X_{ij'}^2(s)+X_{ij'}^3(s)\right]\right\}\times\\ \left[X_{ij'}(s)+X_{ij'}^2(s)+X_{ij'}^3(s)\right]-6s\be_{j'}(s)g^*(s,X_{ij'})-s^2X_{ij'}(s)g^*(s,X_{ij'}),
\end{multline*}
for all $j'=1,...,J$. In this case our $k^{th}$ coefficient would be,
\begin{multline*}
\be_k=\frac{1}{2\left\{\sum_{i=1}^N\left[X_{ik}(s)+X_{ik}^2(s)+X_{ik}^3(s)\right]^2\right\}}\times\\\biggr\{2\sum_{i=1}^NY_i(s)\left[X_{ij'}(s)+X_{ij'}^2(s)+X_{ij'}^3(s)\right]\\-2\sum_{i=1}^N\sum_{j'=0}^{J-1}\be_{j'}(s)\left[X_{ij'}(s)+X_{ij'}^2(s)+X_{ij'}^3(s)\right]^2-s^2X_{ij'}(s)g(s,X_{ij'})\biggr\},
\end{multline*}
where $\sum_{i=1}^N\left[X_{ik}(s)+X_{ik}^2(s)+X_{ik}^3(s)\right]\neq 0$.
\end{example}

\section{Proofs}

\subsection{Proof of Lemma \ref{l0.0}}
Without loss of generality assume absolute value of the quantum Lagrangian 	$|\hat f|\leq 1$ and $||\rho_n||\leq 1$, $||\rho||\leq 1$. Suppose, $\gamma_{\mathcal{T}}$ and $\gamma_{\Omega}$ denote the projections on $\mathcal{T}$ and the sample space $\Omega$, respectively. As $\mathcal{T}$ is a completely regular space, there exists a compact set $\kappa\subset \mathcal{T}$ such that for any $\epsilon>0$ and for all $n$ we have that,
\begin{equation*}
|\rho_n|\circ \gamma_{\mathcal{T}}^{-1}(\mathcal{T}\setminus\kappa)+|\rho|\circ \gamma_{\mathcal{T}}^{-1}(\mathcal{T}\setminus\kappa)\leq\epsilon.
\end{equation*}
The space $K(\kappa)$ is separable because $\kappa$ is Euclidean metrizable. For every $\omega\in\Omega$, define a $g_\omega$ as a continuous function $s,\mathbf{X}\mapsto\hat f(\omega,s,\mathbf X)$ on $\kappa$. Hence, the mapping $g:\Omega\to K(\kappa)$ is Borel. As the projections of measures on $\Omega$ are uniformly countably additive, there exists a probability measure $\theta$ on $\mathcal P$ with respect to which they have uniformly integral densities. By separability of $K(\kappa)$ and applying Lusin's theorem to the mapping $g$ and measure $\theta$, there is a finite partition of $\Omega$ into sets $\mathcal{P}_1,...,\mathcal{P}_p,\mathcal{P}_{p+1}\in\mathcal{P}$ and functions $\hat f_1,...,\hat f_p\in K(\kappa)$ such that $||\E_s\hat f_i||_{K(\kappa)}\leq 1$, $||\E_s g_\omega-\E_s\hat f_i||_{K(\kappa)}\leq\epsilon$ for all $\omega\in\mathcal{P}_i$, $i\leq p$, and
\begin{equation*}
|\rho_n|\circ \gamma_{\Omega}^{-1}(\mathcal{P}_{p+1})+|\rho|\circ \gamma_{\Omega}^{-1}(\mathcal{P}_{p+1})\leq\epsilon, \forall n.
\end{equation*}
As $\mathcal{T}$ is completely regular, every conditional expectation $\E_s\hat f_i$ extends to $\mathcal{T}$ with the preservation of the maximum of the absolute value. By assumption, there exists a time interval index $n_0$ such that the absolute value of the difference between conditional expected integrals of $\E_s h(\omega,s,\mathbf{X}):=\sum_{i=1}^p\mathcal{I}_{\mathcal P_i}(\omega)\E_s\hat f_i(s,\mathbf{X}) $ against the measure $\rho_n$ and $\rho$ does not exceed $\epsilon$ for all $n\geq n_0$, where $\mathcal{I}_{\mathcal P_i}(\omega)$ is the indicator function on partition $\mathcal{P}_i$ on $\mathcal{P}$ \cite{bogachev}. Furthermore, $\sup_{\mathbf X}|\E_s\hat f(\omega,s,\mathbf{X})-E_sh(\omega,s,\mathbf{X})|\leq 2$, $|\E_s\hat f(\omega,s\mathbf{X})-h(\omega,s,\mathbf{X})|\leq\epsilon$ on $\cup_{i=1}^p\mathcal{P}_i\times\kappa$ with
\begin{equation*}
|\rho_n|\left(\Omega\times(\mathcal{T}\setminus\kappa)\right)+|\rho|(\mathcal P_{p+1}\times\mathcal{T})\leq\epsilon.
\end{equation*}
It remains to use the estimate
\begin{equation*}
\int_{\Omega\times\mathcal T}|\E_s\hat f-\E_s h|d|\rho_n|\leq\int_{\cup_{i=1}^p\mathcal P_i\times\kappa}|\E_s\hat f-\E_s h|d|\rho_n|+4\epsilon\leq 5\epsilon,
\end{equation*}
and a similar estimate for $\rho$. Therefore, for $[s,\tau]$ the Equation (\ref{0.1}) holds.

\subsection{Proof of Lemma \ref{l0.1}}
(i). Assumption \ref{as0.1} tells us for small time interval $[s,s+\epsilon]$, $\tilde f\leq\tilde g$ and $\tilde f,\tilde g\geq 0$ in $\mathcal F_s$ two increasing sequences $\{\tilde f_{k_1}\}_{k_1\geq 0}$ and $\{\tilde g_{k_2}\}_{k_2\geq 0}$ such that $\lim_{k_1\ra\infty}\tilde f_{k_1}\leq\lim_{k_2\ra\infty}\tilde g_{k_2}$, hence $\lim_{k_1\ra\infty}\Psi_{s,s+\epsilon}^{k_1}\leq \lim_{k_2\ra\infty}\Psi_{s,s+\epsilon}^{k_2}$, where
\begin{equation*}
\Psi_{s,s+\epsilon}^{k_1}=\frac{1}{N_s^{\tilde f_{k_1}}}\int_{\mathbb R^{N\times J}}\tilde f_{k_1}d\mathbf X
\end{equation*}
and,
\begin{equation*}
\Psi_{s,s+\epsilon}^{k_2}=\frac{1}{N_s^{\tilde g_{k_2}}}\int_{\mathbb R^{N\times J}}\tilde g_{k_2}d\mathbf X.
\end{equation*}
As $\tilde f_{k_1}\leq\lim_{k_2\ra\infty}\tilde g_{k_2}$, the function $\min(\tilde f_{k_1},\tilde g_{k_2})\in\mathcal F_s$ is increasing to $\tilde f_{k_1}$ as $k_2\ra\infty$. Which implies,
\begin{equation*}
\Psi_{s,s+\epsilon}^{k_1}=\lim_{k_2\ra\infty}\frac{1}{N_s^{\min(\tilde f_{k_1},\tilde g_{k_2})}}\int_{\mathbb R^{N\times J}}\min(\tilde f_{k_1},\tilde g_{k_2})d\mathbf X\leq\lim_{k_2\ra\infty}\Psi_{s,s+\epsilon}^{k_2}.
\end{equation*}
From the above condition we know that, for $\epsilon\ra 0$, the transition function $\Psi_{s,s+\epsilon}\in\mathcal F_s^+$ is independent of the choices of increasing sequences converge in $\mathcal F_s^+$ which makes this well defined. Hence, the functionals on $\mathcal F_s^+\cap\mathcal F_s$ coincides with initial functionals and conditions $1$ and $2$ of Assumption \ref{as0.1} hold. If $\tilde f_k$ and $\tilde g_k$ are non-negative in $\mathcal F_s$ and these sequences are increasing to $\tilde f$ and $\tilde g$ then, we have two monotonic limits as $\max(\tilde f,\tilde g)=\lim_{k\ra\infty}\max(\tilde f_k,\tilde g_k)$ and $\min(\tilde f,\tilde g)=\lim_{k\ra\infty}\min(\tilde f_k,\tilde g_k)$. Condition $3$ of Assumption \ref{as0.1} implies $\min(\tilde f,\tilde g)+\max(\tilde f,\tilde g)=\tilde f+\tilde g$ as we assume $\tilde f\leq\tilde g$. Now consider, the sequence $\tilde f_{k_1,k_2}\geq 0$ defined on $\mathcal F_s$ are increasing to $\tilde f_{k_2}\in\mathcal F_s^+$ as $k_1\ra\infty$. Define $\tilde g_{k_3}:=\max_{k_2\leq k_3}\tilde f_{k_2,k_3}$ such that $\tilde g_{k_3}\in\mathcal F_s$. Therefore, as $\tilde g_{k_3}$ is an increasing sequence and for each $k_2\leq k_3$ we have, $\tilde g_{k_3}\leq\tilde g_{k_3+1}$ and $\tilde f_{k_2,k_3}\leq\tilde g_{k_3}\leq\tilde f_{k_3}$. This implies 
\begin{equation*}
\frac{1}{N_s^{\tilde g_{k_3}}}\int_{\mathbb R^{N\times J}}\tilde g_{k_3}d\mathbf X\leq \frac{1}{N_s^{\tilde g_{k_3+1}}}\int_{\mathbb R^{N\times J}}\tilde g_{k_3+1}d\mathbf X
\end{equation*}
and,
\begin{equation*}
\frac{1}{N_s^{\tilde f_{k_2,k_3}}}\int_{\mathbb R^{N\times J}}\tilde f_{k_2,k_3}d\mathbf X\leq \frac{1}{N_s^{\tilde g_{k_3}}}\int_{\mathbb R^{N\times J}}\tilde g_{k_3}d\mathbf X\leq \frac{1}{N_s^{\tilde f_{k_3}}}\int_{\mathbb R^{N\times J}}\tilde f_{k_3}d\mathbf X,
\end{equation*}
as $k_2\leq k_3$. Hence, $\lim_{k_3\ra\infty}\tilde f_{k_3}=\lim_{k_3\ra\infty}\tilde g_{k_3}\in\mathcal F_s^+$ and
\begin{multline*}
\lim_{k_3\ra\infty} \frac{1}{N_s^{\tilde f_{k_3}}}\int_{\mathbb R^{N\times J}}\tilde f_{k_3}d\mathbf X=\lim_{k_3\ra\infty}\frac{1}{N_s^{\tilde g_{k_3}}}\int_{\mathbb R^{N\times J}}\tilde g_{k_3}d\mathbf X\\=\frac{1}{N_s^{\lim_{k_3\ra\infty}\tilde g_{k_3}}}\int_{\mathbb R^{N\times J}}\left[\lim_{k_3\ra\infty}\tilde g_{k_3}\right]d\mathbf X=\frac{1}{N_s^{\lim_{k_3\ra\infty}\tilde f_{k_3}}}\int_{\mathbb R^{N\times J}}\left[\lim_{k_3\ra\infty}\tilde f_{k_3}\right]d\mathbf X.
\end{multline*}
Therefore, Condition $4$ of Assumption \ref{as0.1} is satisfied.

(ii). Define $E$ as subset of $\mathbb{R}^{N\times J}\times\Omega$ such that the indicator function of this set for $[s,s+\epsilon]$ is $\mathcal I_E\in \mathbb{R}^{N\times J}\times\Omega$ such that any function operating in $E$ is on $\mathcal F_s^+$. Now for all $E\subseteq\mathbb R^{N\times J}\times \Omega$ set $\mathbf X(E)=\Psi_{s,s+\epsilon}(\mathcal I_E)$. As $\mathcal I_E\in\mathcal F_s^+$, Condition $3$ in Assumption \ref{as0.1} holds and for two partitions $E_1,E_2\subset E$ we have that, $\mathcal I_{E_1\cap E_1}=\min(\mathcal I_{E_1},\mathcal I_{E_2})$ and $\mathcal I_{E_1\cup E_1}=\max(\mathcal I_{E_1},\mathcal I_{E_2})$. This implies $E$ is closed with respect to finite unions and intersections. Furthermore, by Condition $4$ in Assumption \ref{as0.1} we can say $E$ is closed with respect to countable unions. As we assume $\mathbf X$ is a non-negative monotone additive function hence,
\begin{equation*}
\mathbf X(E_1\cap E_2)+\mathbf X(E_1\cup E_2)=\mathbf X(E_1)+\mathbf X(E_2),
\end{equation*}
such that $\mathbf X=\lim_{k\ra\infty}\mathbf X(E_k)$ for all monotonically increasing sequences of sets $E_k\in E$. Hence, there exists a function 
\begin{equation*}
\mathbf X^*(G)=\inf \{\mathbf X(E):\ E\subset \mathbb R^{N\times J}\times\Omega, G\subset E\}
\end{equation*}
which is countably measurable on Riemann class,
\begin{equation*}
\mathcal A=\left\{A\subset \mathbb R^{N\times J}:\ \mathbf X^*(A)+\mathbf X^*(\mathbb R^{N\times J}\setminus A)=\mathbf K,\ \mathbf K\geq 0\right\}
\end{equation*}
and on the Borel class on filtration $\mathcal F_s^{\mathbf X}$,
\begin{equation*}
\mathcal B=\left\{B\subset\Omega:\ \mathbf X^*(B)+\mathbf X^*(\Omega\setminus B)=1\right\}.
\end{equation*}
Define $\mathbf X$ as the restriction of $\mathbf X^*$ to both $\mathcal A$ and $\mathcal B$.

(iii). Suppose, a set $A^*\subset\mathcal A\times\mathcal B$. As $\tilde f\in\mathcal F_s^+$, then for all constants $c$ we have $\{\tilde f> c\}\in\mathbb R^{N\times J}\times \Omega$, since
\begin{equation*}
\mathcal I_{[\tilde f>c]}=\lim_{k\ra\infty}\min\left[1,k\max(\tilde f-c,0)\right].
\end{equation*}
Therefore, all functions in $\mathcal F_s^+$ are measurable with respect to the $\sigma$-algebra $\sigma(\mathbb R^{N\times J}\times \Omega)$. As we assumed $E\subset\mathbb R^{N\times J}\times \Omega$, there exist an increasing sequence of non-negative functions $\tilde f_k\in \mathcal F_s$ such that $\mathcal I_{E}=\lim_{k\ra \infty}\tilde f_k$ and $\mathbf X^*(E)=\mathbf X(E)=\lim_{k\ra \infty}\Psi_{s,s+\epsilon}^k$. Since $\mathbf X^*(E)+\mathbf X^*(\{\mathbb R^{N\times J}\times\Omega\}\setminus E)\geq 1$, it is sufficient to prove that, $\mathbf X^*(E)+\mathbf X^*(\{\mathbb R^{N\times J}\times\Omega\}\setminus E)\leq 1$ to show $E\in\mathcal A\times\mathcal B$. Hence, it is equivalent to prove
\begin{equation}\label{1}
\mathbf X^*(\{\mathbb R^{N\times J}\times\Omega\}\setminus E)\leq\lim_{k\ra \infty}\Psi_{s,s+\epsilon}(1-\tilde f_k).
\end{equation}
As $\tilde f_k$'s are increasing sequences, $1-\tilde f_k$ are decreasing in $\mathcal I_{\{\mathbb R^{N\times J}\times\Omega\}\setminus E}$. The positive, finite constant $c\in (0,1)\times\mathbb R^{N\times J}$ define a set $\mathcal E=\{1-\tilde f_k>c\}$ contains $\{\mathbb R^{N\times J}\times\Omega\}\setminus E$ and $\mathcal E\subset \mathbb R^{N\times J}\times\Omega$. Hence, the interval on this new indicator function $\mathcal I_{\mathcal E}\leq c^{-1}(1-\tilde f_k)$ implies
\begin{equation*}
\mathbf X^*(\{\mathbb R^{N\times J}\times\Omega\}\setminus E)\leq\mathbf X(\mathcal E)\leq c^{-1}\Psi_{s,s+\epsilon}(1-\tilde f_k),
\end{equation*}
where constant matrix $c^{-1}$ has each element inverted in it. After keeping the space $\mathbb R^{N\times J}$ fixed and letting $c\ra 1$ and $k\ra\infty$ Inequality (\ref{1}) is obtained.

(iv). It is important to know that, all the functions in $\mathcal F_s^+$ are $\mathcal F_s$-measurable. For $E\subset \mathbb R^{N\times J}\times \Omega$ if $\tilde f=\mathcal I_E$, then 
\begin{equation}\label{3}
\Psi_{s,s+\epsilon}(\mathbf X)=\frac{1}{N_s}\int_{\mathbb R^{N\times J}}\tilde f d\mathbf X
\end{equation} 
is satisfied by the way $\mathbf X$ is defined. Furthermore, Equation (\ref{3}) holds for any finite linear combinations of indicators of sets in $\mathbb R^{N\times J}\times \Omega$. Suppose, a non-negative function $\tilde f\in\mathcal F_s^+$ and $\tilde f\leq 1$. Then for any $k\in\mathbb N$, we have that
\begin{equation*}
\tilde f_k:=\sum_{i=1}^{2^k-1}i2^{-k}\mathcal I_{[i2^{-k}<\tilde f<(i+1)2^{-k}]}=2^{-k}\sum_{i=1}^{2^{-k}-1}\mathcal I_{[\tilde f>i2^{-k}]},
\end{equation*}
which follows
\begin{equation*}
\Psi_{s,s+\epsilon}^k(\mathbf X)=\frac{1}{N_s^{\tilde f_k}}\int_{\mathbb R^{N\times J}}\tilde f_k d\mathbf X.
\end{equation*} 
From Conditions $1-4$ in Assumption \ref{as0.1} we know as $k\ra\infty$, the left and right hand sides of the above equality converges to $\Psi_{s,s+\epsilon}(\mathbf X)$ and $\frac{1}{N_s}\int_{\mathbb R^{N\times J}}\tilde fd\mathbf X$ respectively. Moreover, as $\tilde f=\lim_{k\ra\infty}\min(\tilde f,k)$ and $\min(\tilde f,k)\in\mathcal F_s^+$ for all $\tilde f\geq 0$, Equation (\ref{3}) still holds. Finally, for any $\tilde f\in\mathcal F_s$, condition $\tilde f=\max(\tilde f,0)-\max(-\tilde f,0)$ holds and the uniqueness of $\mathbf X$ comes from the fact that $ E$ is closed with respect to finite intersections and it generates a $\sigma$-algebra.
	
\subsection{Proof of Proposition \ref{p0}}
Using Equations (\ref{2}) and (\ref{run}), with initial condition $\mathbf {X}_0$, the Lagrangian of this system is,
\begin{multline*}
\mathcal{L}_{0,T}(\mathbf{X})=
\int_0^{T}\E_{s}\left\{
\sum_{i=1}^{N}\left[Y_i(s)-\sum_{j'=1}^J\beta_{j'}(s)X_{ij'}(s)\right]^2ds\right. \\
+\lambda[{\bf U}(s+ds)-{\bf U}(s)-
\bm{\mu}[s,\bm\be(s),\mathbf{X}(s)]ds\\
\left.\phantom{\int}
-\bm{\sigma}[s,\bm\be(s),\mathbf{X}(s)]d\mathbf{B}(s)]\right\},
\end{multline*}
where $\lambda$ is the time independent Lagrange multiplier which is assumed to be non-negative. Subdivide $[0,T]$ into $n$ equal time-intervals $[s,s+\epsilon]$. For any positive $\epsilon$ and normalizing constant $N_s>0$, define a transition function as 
\begin{equation}\label{3.0}
\Psi_{s,s+\epsilon}(\mathbf{X})=
\frac{1}{N_s}\int_{\mathbb{R}^{N\times J}}\exp
\left[-\epsilon \mathcal{L}_{s,s+\epsilon}(\mathbf{X})\right]
\Psi_s(\mathbf{X})d\mathbf{X},
\end{equation}
where $\Psi_s(\mathbf{X})$ is the transition function at the beginning of $s$ and $\frac{1}{N_s}\ d\mathbf{X}$ is a finite Riemann measure  such that for $k^{th}$ time interval the transition function is,
\begin{equation}\label{3.2}
\Psi_{0,T}(\mathbf{X})=\frac{1}{N_s^n}
\int_{\mathbb{R}^{N\times J\times n}}\exp\left[
-\epsilon\sum_{k=1}^n\mathcal{L}_{s,s+\epsilon}^k(\mathbf{X})\right]
\Psi_0(\mathbf{X})\prod_{k=1}^nd\mathbf{X}^k,
\end{equation}
with the finite measure $N_s^{-n}\prod_{k=1}^nd\mathbf{X}^k$ 
and initial transition function $\Psi_0(\mathbf{X})>0$ for all $n\in\mathbb{N}$ \cite{fujiwara2017}. Equations (\ref{3.0}) and (\ref{3.2}) consider all continuous infinite paths of transition of $\mathbf X$ in any two time intervals.  

Define $\Delta {\bf U}(\nu)={\bf U}(\nu+d\nu)-{\bf U}(\nu)$, then Fubuni's theorem implies,
\begin{multline*}
\mathcal{L}_{s,\tau}(\mathbf{X})=\E_{s}\ \int_s^{\tau}\left\{\sum_{i=1}^{N}\left[Y_i(\nu)-\sum_{j'=1}^J\beta_{j'}(\nu)X_{ij'}(\nu)\right]^2d\nu\right. \\
\left.
\phantom{\int}
+\lambda[\Delta\bf U(\nu)-
\bm{\mu}[\nu,\bm\be(\nu),\mathbf{X}(\nu)]d\nu-
\bm{\sigma}[\nu,\bm\be(\nu),\mathbf{X}(\nu)]d\mathbf{B}(\nu)]
\right\},
\end{multline*}
where $\tau=s+\epsilon$. As we assume the coefficient dynamics has drift and diffusion parts, $\mathbf{X} (\nu)$ is an It\^o process, there exists a smooth function $g[\nu,\mathbf{X}(\nu)]\in C^2([0,T]\times\mathbb{R}^{N\times J})$ such that $\mathbf{I}(\nu)=g[\nu,\mathbf{X}(\nu)]$ where $\mathbf{I}(\nu)$ is an It\^o process  \cite{oksendal2003}. Assuming 
\begin{multline*}
g[\nu+\Delta \nu,\mathbf{X}(\nu)+\Delta \mathbf{X}(\nu)]=\\ 
\lambda[\Delta\bf U(\nu)-\bm\mu[\nu,\bm\be(\nu),\mathbf{X}(\nu)]d\nu-
\bm\sigma[\nu,\bm\be(\nu),\mathbf{X}(\nu)] d\mathbf{B}(\nu)],
\end{multline*}
for a very small time interval around $s$ with $\epsilon\downarrow 0$, generalized It\^o's Lemma yields,
\begin{eqnarray*}
\epsilon\mathcal{L}_{s,\tau}(\mathbf{X}) & = &  
\E_{s}\left\{\epsilon\sum_{i=1}^{N}\left[Y_i(s)-\sum_{j'=1}^J\beta_{j'}(s)X_{ij'}(s)\right]^2+\epsilon g[s,\mathbf{X}(s)] \right.\\
& & +\epsilon g_s[s,\mathbf{X}(s)]+\epsilon 
g_{\mathbf{X}}[s,\mathbf{X}(s)]\bm\mu[s,\bm\be(s),\mathbf{X}(s)] \\
& & 
+\epsilon g_{\mathbf{X}}[s,\mathbf{X}(s)]
\bm{\sigma}[s,\bm\be(s),\mathbf{X}(s)]
\Delta\mathbf{B}(s)\\ 
& & \left.+\mbox{$\frac{1}{2}$} 
\sum_{i=1}^N\sum_{j=1}^N\epsilon
\bm{\sigma}^{ij}[s,\bm\be(s),\mathbf{X}(s)]
g_{X_iX_j}[s,\mathbf{X}(s)]+o(\epsilon)\right\},
\end{eqnarray*}
where $\bm\sigma^{ij}[s,\mathbf{X}(s)]$ represents $\{i,j\}^{th}$ component of the variance-covarience matrix, $g_s=\partial g/\partial s$, $g_{\mathbf{X}}=\partial g/\partial \mathbf{X}$ and $g_{X_iX_j}=\partial^2 g/(\partial X_{ij'}\ \partial X_{jj'})$, $\Delta B_i\ \Delta B_j=\delta^{ij}\ \epsilon$, $\Delta B_i\ \epsilon=\epsilon\ \Delta B_i=0$, and $\Delta X_i(s)\ \Delta X_j(s)=\epsilon$, where $\delta^{ij}$ is the Kronecker delta function. As $\E_s[\Delta \mathbf{B}(s)]=0$ and $\E_s[o(\epsilon)]/\epsilon\ra 0$, for $\epsilon\ra 0$, with the vector of initial conditions $\mathbf{X}_{0_{N\times 1}}$ dividing throughout by  $\epsilon$ and taking the conditional expectation we get,
\begin{eqnarray*}
	\mathcal{L}_{s,\tau}(\mathbf{X}) & = &
	\sum_{i=1}^{N}\left[Y_i(s)-\sum_{j'=1}^J\beta_{j'}(s)X_{ij'}(s)\right]^2
	+g[s,\mathbf{X}(s)] \\ 
	& & +g_s[s,\mathbf{X}(s)]
	+ g_{\mathbf{X}}[s,\mathbf{X}(s)]\bm{\mu}[s,\bm\be(s),\mathbf{X}(s)] \\
	& & +\mbox{$\frac{1}{2}$}\sum_{i=1}^I\sum_{j=1}^I
	\bm{\sigma}^{ij}[s,\bm\be(s),\mathbf{X}(s)]
	g_{X_iX_j}[s,\mathbf{X}(s)]+o(1).
\end{eqnarray*}
Suppose, there exists a vector $\mathbf{\xi}_{N\times 1}$  such that $\mathbf{X}(s)_{N\times 1}=\mathbf{X}(\tau)_{N\times 1}+\xi_{N\times 1}$. For a number $0<\eta<\infty$ assume $|\xi|\leq\eta\epsilon [\mathbf{X}^T(s)]^{-1}$, which makes $\xi$ a very small number for each $\epsilon\downarrow 0$. Furthermore, as $d\xi$ is a cylindrical measure,
\begin{multline}\label{7}
\Psi_s^\tau(\mathbf{X})+\epsilon\frac{\partial \Psi_s^\tau(\mathbf{X})}{\partial s}
+o(\epsilon)=
\frac{1}{N_s}\int_{\mathbb{R}^{N\times J}}
\left[\Psi_s^\tau(\mathbf{X})+\xi\frac{\partial \Psi_s^\tau(\mathbf{X})}{\partial \mathbf{X}}+o(\epsilon)\right]\times \\
\exp\left\{-\epsilon\left[\sum_{i=1}^{N}\left[Y_i(s)-\sum_{j'=1}^J\beta_{j'}(s)[X_{ij'}(\tau)+\xi]\right]^2\right.\right.\\ 
+g[s,\mathbf{X}(\tau)+\xi]+ g_s[s,\mathbf{X}(\tau)+\xi]\\
+g_{\mathbf{X}}[s,\mathbf{X}(\tau)+\xi]\bm{\mu}[s,\bm\be(s),\mathbf{X}(\tau)+\xi]\\ 
\left.\left.+\mbox{$\frac{1}{2}$}\sum_{i=1}^N\sum_{j=1}^N
\bm{\sigma}^{ij}[s,\bm\be(s),\mathbf{X}(\tau)+\xi]
g_{X_iX_j}[s,\mathbf{X}(\tau)+\xi]\right]\right\}d\xi+o(\epsilon^{1/2}).
\end{multline}
After defining a $C^2$ function 
\begin{eqnarray*}
	f[s,\bm\be(s),\xi] & = & 
	\sum_{i=1}^{N}\left[Y_i(s)-\sum_{j'=1}^J\beta_{j'}(s)[X_{ij'}(\tau)+\xi]\right]^2\\ 
	& & +g[s,\mathbf{X}(\tau)+\xi]+g_s[s,\mathbf{X}(\tau)+\xi] \\
	& & +g_{\mathbf{X}}[s,\mathbf{X}(\tau)+\xi]
	\bm{\mu}[s,\bm\be(s),\mathbf{X}(\tau)+\xi]\\
	& & +\mbox{$\frac{1}{2}$}\sum_{i=1}^N\sum_{j=1}^N
	\bm{\sigma}^{ij}[s,\bm\be(s),\mathbf{X}(\tau)+\xi]\times \\
	& & g_{X_iX_j}[s,\mathbf{X}(\tau)+\xi],
\end{eqnarray*} 
Equation (\ref{7}) becomes,
\begin{multline}\label{8}
\Psi_s^\tau(\mathbf{X})+
\epsilon\frac{\partial \Psi_s^\tau(\mathbf{X})}{\partial s} =
\frac{1}{N_s}\Psi_s^\tau(\mathbf{X}) 
\int_{\mathbb{R}^{N\times J}}\exp\left\{-\epsilon f[s,\bm\be(s),\xi]\right\}d\xi\\
+\frac{1}{N_s}\frac{\partial \Psi_s^\tau(\mathbf{X})}{\partial \mathbf{X}}
\int_{\mathbb{R}^{N\times J}} 
\xi\exp\left\{-\epsilon f[s,\bm\be(s),\xi]\right\} d\xi
+o(\epsilon^{1/2}).
\end{multline}
For $\epsilon\downarrow 0$, $\Delta \mathbf{X}\downarrow 0$ and
\begin{multline}\label{9}
f[s,\bm\be(s),\xi]=f[s,\bm\be(s),\mathbf{X}(\tau)]+
\sum_{i=1}^{N}f_{X_i}[s,\bm\be(s),\mathbf{X}(\tau)][\xi_{ij'}-X_{ij'}(\tau)]\\
+\mbox{$\frac{1}{2}$}\sum_{i=1}^{N}\sum_{j=1}^{N}
f_{X_iX_j}[s,\bm\be(s),\mathbf{X}(\tau)]
[\xi_{ij'}-X_{ij'}(\tau)][\xi_{jj'}-X_{jj'}(\tau)]+o(\epsilon).
\end{multline}
We assume there exists a symmetric, positive definite and non-singular Hessian matrix $\mathbf{\Theta}_{(N\times J)\times (N\times J)}$ and a vector $\mathbf{R}_{(N\times J)\times 1}$ such that,
\begin{multline}\label{10}
\int_{\mathbb{R}^{N\times J}} \exp\{-\epsilon f[s,\bm\be(s),\xi]\}d\xi=\\
\sqrt{\frac{(2\pi)^{N\times J}}{\epsilon |\mathbf{\Theta}|}}
\exp\{-\epsilon f[s,\bm\be(s),\mathbf{X}(\tau)]+
\mbox{$\frac{1}{2}$} 
\epsilon\mathbf{R}^T\mathbf{\Theta}^{-1}\mathbf{R}\}.
\end{multline}
The  second Gaussian integral on the right hand side of Equation (\ref{8}) becomes,
\begin{multline}\label{11}
\int_{\mathbb{R}^{N\times J}}\xi\exp\{-\epsilon f[s,\bm\be(s),\xi]\}d\xi=
\sqrt{\frac{(2\pi)^{N\times J}}{\epsilon |\mathbf{\Theta}|}}
\exp\{-\epsilon f[s,\bm\be(s),\mathbf{X}(\tau)]\\
+\mbox{$\frac{1}{2}$}\epsilon\mathbf{R}^T\mathbf{\Theta}^{-1}\mathbf{R}\}
[\mathbf{X}(\tau)+\mbox{$\frac{1}{2}$} (\mathbf{\Theta}^{-1}\ \mathbf{R})].
\end{multline}
Equations (\ref{9}), (\ref{10}) and (\ref{11}) imply
\begin{multline*}
\Psi_s^\tau(\mathbf{X})+\epsilon\frac{\partial \Psi_s^\tau(\mathbf{X})}{\partial s}
=\frac{1}{N_s}\sqrt{\frac{(2\pi)^{N\times J}}{\epsilon |\mathbf{\Theta}|}}
\exp\{-\epsilon f[s,\bm\be(s),\mathbf{X}(\tau)]+
\mbox{$\frac{1}{2}$}\epsilon\mathbf{R}^T\mathbf{\Theta}^{-1}\mathbf{R}\}\\
\times
\left\{\Psi_s^\tau(\mathbf{X})+[\mathbf{X}(\tau)+\mbox{$\frac{1}{2}$}
(\mathbf{\Theta}^{-1}\mathbf{R})]
\frac{\partial \Psi_{\mathbf{s}}^\tau(\mathbf{X})}{\partial \mathbf{X}}\right\}
+o(\epsilon^{1/2}).
\end{multline*}
Assuming $N_s=\sqrt{(2\pi)^{N\times J}/(\epsilon |\mathbf{\Theta}|)}>0$, we get Wick rotated 
Schr\"odinger type equation as,
\begin{multline}\label{13}
\Psi_s^\tau(\mathbf{X})+\epsilon\frac{\partial \Psi_s^\tau(\mathbf{X})}{\partial s}= 
\{1-\epsilon f[s,\bm\be(s),\mathbf{X}(\tau)]+
\mbox{$\frac{1}{2}$}\epsilon\mathbf{R}^T\mathbf{\Theta}^{-1}\mathbf{R}\}\times \\
\left\{\Psi_s^\tau(\mathbf{X})+[\mathbf{X}(\tau)+\mbox{$\frac{1}{2}$} 
(\mathbf{\Theta}^{-1}\mathbf{R})]
\frac{\partial \Psi_{\mathbf{s}}^\tau(\mathbf{X})}{\partial \mathbf{X}}\right\}
+o(\epsilon^{1/2}).
\end{multline}
For any finite positive number $\eta$ we know $\mathbf{X}(\tau)\leq\eta\epsilon|\xi^T|^{-1}$. Then there exists $|\mathbf{\Theta}^{-1}\mathbf{R}|\leq 2 \eta\epsilon|1-\xi^T|^{-1}$ such that for $\epsilon\downarrow 0$ we have, $\big|\mathbf{X}(\tau)+\mbox{$\frac{1}{2}$}\ \left(\mathbf{\Theta}^{-1}\ \mathbf{R}\right)\big|\leq\eta\epsilon$ and Equation (\ref{13}) becomes,
\begin{equation}\label{14}
\frac{\partial \Psi_s^\tau(\mathbf{X})}{\partial s}=
\{-f[s,\bm\be(s),\mathbf{X}(\tau)]+
\mbox{$\frac{1}{2}$}\mathbf{R}^T\mathbf{\Theta}^{-1}\mathbf{R}\}
\Psi_s^\tau(\mathbf{X}). 
\end{equation}
As $|\mathbf{\Theta}^{-1}\mathbf{R}|\leq 2 \eta\epsilon|1-\xi^T|^{-1}$, where $\xi^T$ is the transpose of $\xi$, then at $\epsilon\downarrow 0$ we can ignore the second term. Therefore, Equation (\ref{14}) becomes 
\[
\frac{\partial \Psi_s^\tau(\mathbf{X})}{\partial s}=
-f[s,\bm\be(s),\mathbf{X}(\tau)]\Psi_s^\tau(\mathbf{Z}),
\]
and the partial derivative with $\be_{j'}$ yields,
\begin{equation}\label{16}
-\frac{\partial}{\partial \be_{j'}}
f[u,\bm\be(s),\mathbf{X}(\tau)]\Psi_s^\tau(\mathbf{X})=0.
\end{equation}
In Equation (\ref{16}) either $\Psi_s^\tau(\mathbf{X})=0$ or $\frac{\partial }{\partial \be_{j'}}f[s,\bm\be(s),\mathbf{X}(\tau)]=0$. As $\Psi_s^\tau(\mathbf{X})$ is a transition wave function it cannot be zero. Therefore, the partial derivative with respect to $\be_{j'}$ has to be zero. We know, $\mathbf{X}(\tau)=\mathbf{X}(s)-\xi$ and for $\xi\downarrow 0$ as we are looking for some stable solution therefore, in Equation (\ref{16}) $\mathbf{X}(\tau)$ can be replaced by $\mathbf{X}(s)$.
Hence,
\begin{multline}\label{16.1}
f[s,\bm\be(s),\mathbf{X}(s)]=
\sum_{i=1}^{N}\left[Y_i(s)-\sum_{j'=1}^J\beta_{j'}(s)X_{ij'}(s)\right]^2\\
+g[s,\mathbf{X}(s)]+ g_s[s,\mathbf{X}(s)]+
g_{\mathbf{X}}[s,\mathbf{X}(s)]\bm{\mu}[s,\bm\be(s),\mathbf{X}(s)]\\
+\mbox{$\frac{1}{2}$}\sum_{i=1}^N\sum_{j=1}^N
\bm{\sigma}^{ij}[s,\bm\be(s),\mathbf{X}(s)]
g_{X_iX_j}[s,\mathbf{X}(s)].
\end{multline}
Equations (\ref{16}) and (\ref{16.1}) then imply
\begin{multline}\label{16.2}
2\sum_{i=1}^{N}\left[Y_i(s)-\sum_{j'=1}^J\beta_{j'}(s)X_{ij'}(s)\right] X_{ij'}(s)\\
-g_{\mathbf{X}}[s,\mathbf{X}(s)]
\frac{\partial \bm{\mu}[u,\bm{\be}(s),\mathbf{X}(s)]}{\partial \bm{\be}(s)}
\frac{\partial \bm{\be}(s)}{\partial \be_{j'}(s) }\\ 
-\mbox{$\frac{1}{2}$}\sum_{i=1}^N\sum_{j=1}^N
g_{X_iX_j}[s,\mathbf{X}(s)]
\frac{\partial \bm{\sigma}^{ij}[s,\bm{\be}(s),\mathbf{X}(s)]}
{\partial \bm{\be}(s)}
\frac{\partial \bm{\be}(s)}{\partial \be_{j'}(s)}=0.
\end{multline}
Optimal $\be_{j'}(s)$ can be obtained by solving Equation (\ref{16.2}).

\subsection{Proof of Proposition \ref{p1}}
Using Equations (\ref{run}) and (\ref{2.1}), with initial condition $\mathbf {X}_0$ with its basis $h(\mathbf X_0)$, the dynamic Lagrangian action of this system of smoothing spline is,
\begin{multline*}
\mathcal{L}_{0,T}^*(\mathbf{X})=
\int_0^{T}\E_{s}\left\{
\sum_{i=1}^{N}\left[Y_i(s)-\sum_{j'=1}^J\beta_{j'}(s)h[X_{ij'}(s)]\right]^2ds\right. \\
+\lambda^*[{\bf U}(s+ds)-{\bf U}(s)-
\bm{\mu}[s,\bm\be(s),\mathbf{X}(s)]ds\\
\left.\phantom{\int}
-\bm{\sigma}[s,\bm\be(s),\mathbf{X}(s)]d\mathbf{B}(s)]\right\},
\end{multline*}
where $\lambda^*$ is the time independent non-negative penalizing constant. After subdividing $[0,T]$ into $n$ equal time-intervals $[s,s+\epsilon]$ such that for all $\epsilon$ and $N_s^*>0$, define a transition function as 
\begin{equation}\label{4.0}
\Psi_{s,s+\epsilon}^*(\mathbf{X})=
\frac{1}{N_s^*}\int_{\mathbb{R}^{N\times J}}\exp
\left[-\epsilon \mathcal{L}_{s,s+\epsilon}^*(\mathbf{X})\right]
\Psi_s^*(\mathbf{X})d\mathbf{X},
\end{equation}
where $\Psi_s^*(\mathbf{X})$ is the transition function at the beginning of $s$ and $\frac{1}{N_s^*}\ d\mathbf{X}$ is a finite Riemann measure  such that for $k^{th}$ time interval this transition function is,
\begin{equation}\label{4.2}
\Psi_{0,T}^*(\mathbf{X})=\frac{1}{(N_s^*)^n}
\int_{\mathbb{R}^{N\times J\times n}}\exp\left[
-\epsilon\sum_{k=1}^n\mathcal{L}_{s,s+\epsilon}^{k*}(\mathbf{X})\right]
\Psi_0^*(\mathbf{X})\prod_{k=1}^nd\mathbf{X}^k,
\end{equation}
with the finite measure $(N_s^*)^{-n}\prod_{k=1}^nd\mathbf{X}^k$ 
and initial transition function $\Psi_0^*(\mathbf{X})>0$ for all $n\in\mathbb{N}$. Equations (\ref{4.0}) and (\ref{4.2}) consider all continuous infinite paths of transition of $\mathbf X$ in any two time intervals.

Fubuni's theorem implies,
\begin{multline*}
\mathcal{L}_{s,\tau}^*(\mathbf{X})=\E_{s}\ \int_s^{\tau}\left\{\sum_{i=1}^{N}\left[Y_i(\nu)-\sum_{j'=1}^J\beta_{j'}(\nu)h[X_{ij'}(\nu)]\right]^2d\nu\right. \\
\left.
\phantom{\int}
+\lambda^*[\Delta{\bf U}(\nu)-
\bm{\mu}[\nu,\bm\be(\nu),\mathbf{X}(\nu)]d\nu-
\bm{\sigma}[\nu,\bm\be(\nu),\mathbf{X}(\nu)]d\mathbf{B}(\nu)]
\right\},
\end{multline*}
where $\tau=s+\epsilon$ and $\Delta {\bf U}(\nu)={\bf U}(\nu+d\nu)-{\bf U}(\nu)$. As like before the coefficient dynamics has drift and diffusion parts, $\mathbf{X} (\nu)$ is an It\^o process, there exists a smooth function $g^*[\nu,\mathbf{X}(\nu)]\in C^2([0,T]\times\mathbb{R}^{N\times J})$ such that $\mathbf{I}^*(\nu)=g^*[\nu,\mathbf{X}(\nu)]$ where $\mathbf{I}^*(\nu)$ is an It\^o process of the smoothing spline. Assuming 
\begin{multline*}
g^*[\nu+\Delta \nu,\mathbf{X}(\nu)+\Delta \mathbf{X}(\nu)]=\\ 
\lambda^*[\Delta{\bf U}(\nu)-\bm\mu[\nu,\bm\be(\nu),\mathbf{X}(\nu)]d\nu-
\bm\sigma[\nu,\bm\be(\nu),\mathbf{X}(\nu)] d\mathbf{B}(\nu)],
\end{multline*}
for a very small time interval around $s$ with $\epsilon\downarrow 0$, generalized It\^o's Lemma yields,
\begin{eqnarray*}
	\epsilon\mathcal{L}_{s,\tau}^*(\mathbf{X}) & = &  
	\E_{s}\left\{\epsilon\sum_{i=1}^{N}\left[Y_i(s)-\sum_{j'=1}^J\beta_{j'}(s)h[X_{ij'}(s)]\right]^2+\epsilon g^*[s,\mathbf{X}(s)] \right.\\
	& & +\epsilon g_s^*[s,\mathbf{X}(s)]+\epsilon 
	g_{\mathbf{X}}^*[s,\mathbf{X}(s)]\bm\mu[s,\bm\be(s),\mathbf{X}(s)] \\
	& & 
	+\epsilon g_{\mathbf{X}}^*[s,\mathbf{X}(s)]
	\bm{\sigma}[s,\bm\be(s),\mathbf{X}(s)]
	\Delta\mathbf{B}(s)\\ 
	& & \left.+\mbox{$\frac{1}{2}$} 
	\sum_{i=1}^N\sum_{j=1}^N\epsilon
	\bm{\sigma}^{ij}[s,\bm\be(s),\mathbf{X}(s)]
	g_{X_iX_j}^*[s,\mathbf{X}(s)]+o(\epsilon)\right\},
\end{eqnarray*}
where $\bm\sigma^{ij}[s,\mathbf{X}(s)]$ represents $\{i,j\}^{th}$ component of the variance-covarience matrix, $g_s^*=\partial g^*/\partial s$, $g_{\mathbf{X}}^*=\partial g^*/\partial \mathbf{X}$ and $g_{X_iX_j}^*=\partial^2 g^*/(\partial X_{ij'}\ \partial X_{jj'})$, $\Delta B_i\ \Delta B_j=\delta^{ij}\ \epsilon$, $\Delta B_i\ \epsilon=\epsilon\ \Delta B_i=0$, and $\Delta X_i(s)\ \Delta X_j(s)=\epsilon$, where $\delta^{ij}$ is the Kronecker delta function. As $\E_s[\Delta \mathbf{B}(s)]=0$ and $\E_s[o(\epsilon)]/\epsilon\ra 0$, for $\epsilon\downarrow 0$, with the vector of initial conditions $\mathbf{X}_{0_{N\times 1}}$ dividing throughout by  $\epsilon$ and taking the conditional expectation we get,
\begin{eqnarray*}
	\mathcal{L}_{s,\tau}^*(\mathbf{X}) & = &
	\sum_{i=1}^{N}\left[Y_i(s)-\sum_{j'=1}^J\beta_{j'}(s)h[X_{ij'}(s)]\right]^2
	+g^*[s,\mathbf{X}(s)] \\ 
	& & +g_s^*[s,\mathbf{X}(s)]
	+ g_{\mathbf{X}}^*[s,\mathbf{X}(s)]\bm{\mu}[s,\bm\be(s),\mathbf{X}(s)] \\
	& & +\mbox{$\frac{1}{2}$}\sum_{i=1}^I\sum_{j=1}^I
	\bm{\sigma}^{ij}[s,\bm\be(s),\mathbf{X}(s)]
	g_{X_iX_j}^*[s,\mathbf{X}(s)]+o(1).
\end{eqnarray*}
Suppose, there exists a vector $\mathbf{\xi}_{N\times 1}$  such that $\mathbf{X}(s)_{N\times 1}=\mathbf{X}(\tau)_{N\times 1}+\xi_{N\times 1}$. For a number $0<\eta<\infty$ assume $|\xi|\leq\eta\epsilon [\mathbf{X}^T(s)]^{-1}$. Furthermore,
\begin{multline}\label{4.3}
\Psi_s^{\tau*}(\mathbf{X})+\epsilon\frac{\partial \Psi_s^{\tau*}(\mathbf{X})}{\partial s}
+o(\epsilon)=
\frac{1}{N_s^*}\int_{\mathbb{R}^{N\times J}}
\left[\Psi_s^{\tau*}(\mathbf{X})+\xi\frac{\partial \Psi_s^{\tau*}(\mathbf{X})}{\partial \mathbf{X}}+o(\epsilon)\right]\times \\
\exp\left\{-\epsilon\left[\sum_{i=1}^{N}\left[Y_i(s)-\sum_{j'=1}^J\beta_{j'}(s)h[X_{ij'}(\tau)+\xi]\right]^2\right.\right.\\ 
+g^*[s,\mathbf{X}(\tau)+\xi]+ g_s^*[s,\mathbf{X}(\tau)+\xi]\\
+g_{\mathbf{X}}^*[s,\mathbf{X}(\tau)+\xi]\bm{\mu}[s,\bm\be(s),\mathbf{X}(\tau)+\xi]\\ 
\left.\left.+\mbox{$\frac{1}{2}$}\sum_{i=1}^N\sum_{j=1}^N
\bm{\sigma}^{ij}[s,\bm\be(s),\mathbf{X}(\tau)+\xi]
g_{X_iX_j}^*[s,\mathbf{X}(\tau)+\xi]\right]\right\}d\xi+o(\epsilon^{1/2}).
\end{multline}
After defining a $C^2$ function 
\begin{eqnarray*}
	f^*[s,\bm\be(s),\xi] & = & 
	\sum_{i=1}^{N}\left[Y_i(s)-\sum_{j'=1}^J\beta_{j'}(s)h[X_{ij'}(\tau)+\xi]\right]^2\\ 
	& & +g^*[s,\mathbf{X}(\tau)+\xi]+g_s^*[s,\mathbf{X}(\tau)+\xi] \\
	& & +g_{\mathbf{X}}^*[s,\mathbf{X}(\tau)+\xi]
	\bm{\mu}[s,\bm\be(s),\mathbf{X}(\tau)+\xi]\\
	& & +\mbox{$\frac{1}{2}$}\sum_{i=1}^N\sum_{j=1}^N
	\bm{\sigma}^{ij}[s,\bm\be(s),\mathbf{X}(\tau)+\xi]\times \\
	& & g_{X_iX_j}^*[s,\mathbf{X}(\tau)+\xi],
\end{eqnarray*} 
Equation (\ref{4.3}) becomes,
\begin{multline}\label{4.4}
\Psi_s^{\tau*}(\mathbf{X})+
\epsilon\frac{\partial \Psi_s^{\tau*}(\mathbf{X})}{\partial s} =
\frac{1}{N_s^*}\Psi_s^{\tau*}(\mathbf{X}) 
\int_{\mathbb{R}^{N\times J}}\exp\left\{-\epsilon f^*[s,\bm\be(s),\xi]\right\}d\xi\\
+\frac{1}{N_s^*}\frac{\partial \Psi_s^{\tau*}(\mathbf{X})}{\partial \mathbf{X}}
\int_{\mathbb{R}^{N\times J}} 
\xi\exp\left\{-\epsilon f^*[s,\bm\be(s),\xi]\right\} d\xi
+o(\epsilon^{1/2}).
\end{multline}
For $\epsilon\downarrow0$, $\Delta \mathbf{X}\downarrow0$ and
\begin{multline}\label{4.5}
f^*[s,\bm\be(s),\xi]=f^*[s,\bm\be(s),\mathbf{X}(\tau)]+
\sum_{i=1}^{N}f_{X_i}^*[s,\bm\be(s),\mathbf{X}(\tau)][\xi_{ij'}-X_{ij'}(\tau)]\\
+\mbox{$\frac{1}{2}$}\sum_{i=1}^{N}\sum_{j=1}^{N}
f_{X_iX_j}^*[s,\bm\be(s),\mathbf{X}(\tau)]
[\xi_{ij'}-X_{ij'}(\tau)][\xi_{jj'}-X_{jj'}(\tau)]+o(\epsilon).
\end{multline}
We assume there exists a symmetric, positive definite and non-singular Hessian matrix $\mathbf{\Theta}_{(N\times J)\times (N\times J)}$ and a vector $\mathbf{R}_{(N\times J)\times 1}$ such that,
\begin{multline}\label{4.6}
\int_{\mathbb{R}^{N\times J}} \exp\{-\epsilon f^*[s,\bm\be(s),\xi]\}d\xi=\\
\sqrt{\frac{(2\pi)^{N\times J}}{\epsilon |\mathbf{\Theta}|}}
\exp\{-\epsilon f^*[s,\bm\be(s),\mathbf{X}(\tau)]+
\mbox{$\frac{1}{2}$} 
\epsilon\mathbf{R}^T\mathbf{\Theta}^{-1}\mathbf{R}\}.
\end{multline}
The  second Gaussian integral on the right hand side of Equation (\ref{4.4}) becomes,
\begin{multline}\label{4.7}
\int_{\mathbb{R}^{N\times J}}\xi\exp\{-\epsilon f^*[s,\bm\be(s),\xi]\}d\xi=
\sqrt{\frac{(2\pi)^{N\times J}}{\epsilon |\mathbf{\Theta}|}}
\exp\{-\epsilon f^*[s,\bm\be(s),\mathbf{X}(\tau)]\\
+\mbox{$\frac{1}{2}$}\epsilon\mathbf{R}^T\mathbf{\Theta}^{-1}\mathbf{R}\}
[\mathbf{X}(\tau)+\mbox{$\frac{1}{2}$} (\mathbf{\Theta}^{-1}\ \mathbf{R})].
\end{multline}
Equations (\ref{4.5}), (\ref{4.6}) and (\ref{4.7}) imply
\begin{multline*}
\Psi_s^{\tau*}(\mathbf{X})+\epsilon\frac{\partial \Psi_s^{\tau*}(\mathbf{X})}{\partial s}
=\frac{1}{N_s^*}\sqrt{\frac{(2\pi)^{N\times J}}{\epsilon |\mathbf{\Theta}|}}
\exp\{-\epsilon f^*[s,\bm\be(s),\mathbf{X}(\tau)]\\+
\mbox{$\frac{1}{2}$}\epsilon\mathbf{R}^T\mathbf{\Theta}^{-1}\mathbf{R}\}
\left\{\Psi_s^{\tau*}(\mathbf{X})+[\mathbf{X}(\tau)+\mbox{$\frac{1}{2}$}
(\mathbf{\Theta}^{-1}\mathbf{R})]
\frac{\partial \Psi_{\mathbf{s}}^{\tau*}(\mathbf{X})}{\partial \mathbf{X}}\right\}
+o(\epsilon^{1/2}).
\end{multline*}
Assuming $N_s^*=\sqrt{(2\pi)^{N\times J}/(\epsilon |\mathbf{\Theta}|)}>0$, the Wick rotated 
Schr\"odinger type equation is,
\begin{multline}\label{4.8}
\Psi_s^{\tau*}(\mathbf{X})+\epsilon\frac{\partial \Psi_s^{\tau*}(\mathbf{X})}{\partial s}= 
\{1-\epsilon f^*[s,\bm\be(s),\mathbf{X}(\tau)]+
\mbox{$\frac{1}{2}$}\epsilon\mathbf{R}^T\mathbf{\Theta}^{-1}\mathbf{R}\}\times \\
\left\{\Psi_s^{\tau*}(\mathbf{X})+[\mathbf{X}(\tau)+\mbox{$\frac{1}{2}$} 
(\mathbf{\Theta}^{-1}\mathbf{R})]
\frac{\partial \Psi_{\mathbf{s}}^{\tau*}(\mathbf{X})}{\partial \mathbf{X}}\right\}
+o(\epsilon^{1/2}).
\end{multline}
As $\mathbf{X}(\tau)\leq\eta\epsilon|\xi^T|^{-1}$, there exists $|\mathbf{\Theta}^{-1}\mathbf{R}|\leq 2 \eta\epsilon|1-\xi^T|^{-1}$ such that for $\epsilon\downarrow 0$ we have, $\big|\mathbf{X}(\tau)+\mbox{$\frac{1}{2}$}\ \left(\mathbf{\Theta}^{-1}\ \mathbf{R}\right)\big|\leq\eta\epsilon$ and Equation (\ref{4.8}) becomes,
\begin{equation*}
\frac{\partial \Psi_s^{\tau*}(\mathbf{X})}{\partial s}=
\{-f^*[s,\bm\be(s),\mathbf{X}(\tau)]+
\mbox{$\frac{1}{2}$}\mathbf{R}^T\mathbf{\Theta}^{-1}\mathbf{R}\}
\Psi_s^{\tau*}(\mathbf{X}). 
\end{equation*}
As $|\mathbf{\Theta}^{-1}\mathbf{R}|\leq 2 \eta\epsilon|1-\xi^T|^{-1}$, where $\xi^T$ is the transpose of $\xi$, then we have,
\[
\frac{\partial \Psi_s^{\tau*}(\mathbf{X})}{\partial s}=
-f^*[s,\bm\be(s),\mathbf{X}(\tau)]\Psi_s^{\tau*}(\mathbf{Z}),
\]
and the partial derivative with $\be_{j'}$ yields,
\begin{equation}\label{4.9}
-\frac{\partial}{\partial \be_{j'}}
f^*[u,\bm\be(s),\mathbf{X}(\tau)]\Psi_s^{\tau*}(\mathbf{X})=0.
\end{equation}
In Equation (\ref{4.9}) either $\Psi_s^{\tau*}(\mathbf{X})=0$ or $\frac{\partial }{\partial \be_{j'}}f^*[s,\bm\be(s),\mathbf{X}(\tau)]=0$. As $\Psi_s^{\tau*}(\mathbf{X})$ is a transition wave function it cannot be zero. Therefore, the partial derivative with respect to $\be_{j'}$ has to be zero. We know, $\mathbf{X}(\tau)=\mathbf{X}(s)-\xi$ and for $\xi\downarrow 0$ as we are looking for some stable solution therefore, in Equation (\ref{4.9}) $\mathbf{X}(\tau)$ can be replaced by $\mathbf{X}(s)$.
Hence,
\begin{multline}\label{4.10}
f^*[s,\bm\be(s),\mathbf{X}(s)]=
\sum_{i=1}^{N}\left[Y_i(s)-\sum_{j'=1}^J\beta_{j'}(s)h[X_{ij'}(s)]\right]^2\\
+g^*[s,\mathbf{X}(s)]+ g_s[s,\mathbf{X}(s)]+
g_{\mathbf{X}}^*[s,\mathbf{X}(s)]\bm{\mu}[s,\bm\be(s),\mathbf{X}(s)]\\
+\mbox{$\frac{1}{2}$}\sum_{i=1}^N\sum_{j=1}^N
\bm{\sigma}^{ij}[s,\bm\be(s),\mathbf{X}(s)]
g_{X_iX_j}^*[s,\mathbf{X}(s)].
\end{multline}
Equations (\ref{4.9}) and (\ref{4.10}) then imply
\begin{multline}\label{4.11}
2\sum_{i=1}^{N}\left[Y_i(s)-\sum_{j'=1}^J\beta_{j'}(s)h[X_{ij'}(s)]\right] h[X_{ij'}(s)]\\
-g_{\mathbf{X}}^*[s,\mathbf{X}(s)]
\frac{\partial \bm{\mu}[u,\bm{\be}(s),\mathbf{X}(s)]}{\partial \bm{\be}(s)}
\frac{\partial \bm{\be}(s)}{\partial \be_{j'}(s) }\\ 
-\mbox{$\frac{1}{2}$}\sum_{i=1}^N\sum_{j=1}^N
g_{X_iX_j}^*[s,\mathbf{X}(s)]
\frac{\partial \bm{\sigma}^{ij}[s,\bm{\be}(s),\mathbf{X}(s)]}
{\partial \bm{\be}(s)}
\frac{\partial \bm{\be}(s)}{\partial \be_{j'}(s)}=0.
\end{multline}
Optimal $\be_{j'}(s)$ can be obtained by solving Equation (\ref{4.11}).

\section{Discussion}
In Lemmas \ref{l0.0} and \ref{l0.1} we show the existence of path integral in penalized regression. Proposition \ref{p0} helps us determining the coefficients in more generalized LASSO type frameworks. Then we provide seven cases to obtain a closed form $\be_k$, which are functions of $X_{ik}, X_{ij'}, Y_i$ and $\be_{j'}$. Furthermore, in cases like LASSO, standard $L^p$-norm, elastic net regression, fused LASSO and bridge regression we assume $\be_{j'}$'s are non-zero to get rid of the problem of non-differentiability. Proposition \ref{p1} determines optimal $\be$ coefficients under generalized spline environment where $h(X_{ij'})$ represents any time dependent basis function and Example \ref{e8} considers a dynamic cubic smoothing spline. Throughout this paper we assume $g(s,X_{ij'})=\lambda^*\exp(sX_{ij'})$ and diffusion coefficient as $2\sum_{i=1}^N\be_{j'}X_{ij'}$ to make our result comprehensible and hence, $\be_k$'s are easily comparable among our eight examples. In our future research we will extend this idea into more generalized Riemann manifold.
\bibliography{bib}
\end{document}